\documentstyle[preprint,prb,aps,epsfig]{revtex}
\begin{document}

\draft

\title{Collective Modes in Supersolid $^{4}$He}

\author{M.J. Bijlsma and H.T.C. Stoof}

\address{Institute for Theoretical Physics, University of Utrecht,
         Princetonplein 5, P.O. Box 80.006, 3508 TA Utrecht, The Netherlands}

\maketitle

\begin{abstract}
We derive the hydrodynamic equations of motion of solid and
supersolid $^{4}$He, that
describe the collective modes of these phases.
In particular, the usual hydrodynamics is modified in such a way
that it leads to the presence of a propagating instead of a
diffusive defect mode. The former is appropriate
for a quantum crystal and observed in recent experiments.
Furthermore, we find that in supersolid helium there are
two additional modes associated with the superfluid degrees of freedom.
The observation of these additional modes is a clear experimental
signature of the supersolid phase.
\end{abstract}


\section{INTRODUCTION}

The low temperature behavior of the strongly
interacting quantum liquid $^{4}$He has been a subject
of experimental and theoretical research for decades.
In 1908 Helium was first liquified by Kamerlingh Onnes and in 1911
he discovered a sharp maximum in the density at what is now
commonly called
the $\lambda$-point.\cite{Keller}
After that, a number of macroscopic
quantum phenomena like superfluid flow, second sound,
the fountain effect and quantized vortices were observed.
Phenomenological theories were developed and
justified from a microscopic point of view.\cite{Landau2,Feynman,Hohenberg}
Also, the famous Kosterlitz-Thouless transition was first
observed in thin superfluid helium films.\cite{Rudnick}

In the solid phase of $^{4}$He, which
is reached only at low temperature
and high pressure (cf. Fig. \ref{fig1}),
one also expects to observe macroscopic
quantum phenomena because of the large zero point vibration
of the atoms about their equilibrium position.\cite{Varma}
Because of this, solid helium has been termed a quantum solid.

In such a solid, the interstitials and vacancies
are effectively delocalized
due to their ability to tunnel through the potential barriers.
At low temperatures these point defects then
form a weakly interacting Bose gas.
Furthermore, the large zero point motion results in
an unusually rapid exchange rate of nearest neighbour atoms,
which may lead to large ring exchanges between the helium atoms.\cite{Feynman}
Bose-Einsein condensation of the defects or exchange processes
of the lattice atoms may then open two routes to a new phase
of matter at low temperature in which long-range crystalline
and superfluid order coexist. This is called the supersolid phase.
\cite{Meisel}

Theoretically the existence of such a phase has since
long been anticipated.\cite{Andreev,Chester,Leggett2}
However, it was only recently claimed to have
been observed experimentally that three dimensional solid $^{4}$He is
a spatially ordered superfluid, or supersolid, at sufficiently low
temperatures and densities.
The experiments leading to this claim were performed
by Lengua and Goodkind, who measured the attenuation and
velocity of sound in solid $^{4}$He for relatively high
purities and low atomic densities of the quantum crystal.\cite{Lengua}
The temperature dependence of the attenuation revealed a
coupling to thermally activated  excitations, consistent
with the existence of a propagating mode in the gas of point defects
that is expected to be present in a quantum crystal.\cite{Andreev,Stoof}
Furthermore, assuming the speed of sound of the defect mode
to depend on the density of defects in the same way as in a
dilute Bose gas,
they found a relation between the temperature dependence of the phase
velocity and that of the defect density. To consistently interpret their
data they then had to assume a macroscopic
population of the zero momentum state of the point defects, i.e.
a Bose-Einstein condensation of the point defects.
Thus the phase diagram of $^{4}$He in three dimensions
would be qualitatively given by Fig. \ref{fig1}.
Following a certain trajectory in this phase diagram,
$^{4}$He may undergo a transition
from the normal phase to the superfluid phase
at some temperature $T_{\lambda}$ and subsequently from the
superfluid to the supersolid phase at a temperature $T_{c}$.
As mentioned above, the possibility of superfluid flow
in a solid has since long been anticipated theoretically.
Andreev and Lifschitz were the first
to attempt to derive the hydrodynamics of a
supersolid by including the effect of Bose-Einstein condensation
of the defects on the hydrodynamics of an ideal crystal.\cite{Andreev}
In addition to this pioneering work, Liu has more recently
presented a thorough discussion of
the Andreev and Lifschitz hydrodynamics.\cite{Liu}

However, it was pointed out by Martin {\it et.al.}
that the conventional treatment
of the hydrodynamic equations for a classical crystal,
which doesn't include defects, is neccesarily incomplete since
it yields the wrong number of modes.\cite{Martin}
They identified the missing mode
as a mode in the defect density. This implies that the hydrodynamics of
Andreev and Lifshits is also incomplete,
because it does not include the non-condensed defects
and as a result does not lead to the required
defect mode in the normal state of the crystal.
In addition, Martin {\it et.al.} assume diffusive
dynamics for their defect mode. This
seems to be appropriate for a classical but not for a quantum
crystal, where the defect mode is expected to be a propagating mode,
as is confirmed by the experiments of Lengua and Goodkind.

Recently Stoof {\it et.al.}, in respons to experiments
with submonolayer superfluid helium film,\cite{Chen}
derived the hydrodynamic equations
for an isotropic supersolid in two dimensions
which did include propagating behavior of the crucial defect mode.\cite{Stoof}
Moreover, the longitudinal part of the solid hydrodynamics
derived by these authors turns out to be identical to the system
of two coupled wave equations that Lengua and Goodkind used to
accurately model their data.
However, to apply these promising results to the experiments with
solid helium, we have to extend them in two ways. First of all we have
to consider a three dimensional system, and second of all we have
to take into account the anisotropy of solid $^{4}$He, which
is a hexogonally closed packed (HCP) crystal.
Thus we hope to justify from a microscopic point of view
the phenomenological equations that succesfully explained the
propagation of sound in solid $^{4}$He and led to the first claim of
a supersolid phase in  this system.

The paper is organized as follows. In Sec. \ref{section2}
the hydrodynamic equations
describing a normal solid with point defects will be derived.
This is achieved
by deriving an action describing a solid with dislocations,
using methods developed by Kleinert.\cite{Kleinert3}
{}From this action we obtain the interaction between phonons and a point
defect,
 by seeing the point defect
as a limiting case of a dislocation.
Also, a more microscopic point of view is
presented and dissipation is included.
In Sec. \ref{section3} we then add a superfluid degree
of freedom to our hydrodynamic equations in the usual way
and in Sec. \ref{section4} we discuss the
experiment by Lengua and Goodkind in the light of our results.
It should be noted that in order to understand this experiment
it is not neccessary to include temperature flucuations into our
considerations and we will neglect them in the rest of this article.
We conclude with a discussion and outlook in Sec. \ref{section5}.
\section{hydrodynamics of solids with point defects}
\label{section2}
In this section we derive the hydrodynamics of a solid with point
defects. This will be done by first considering the action
describing phonons and their interaction with dislocation loops.
We then obtain
the interaction of phonons with vacancies and interstitials by
shrinking a dislocation loop to zero radius
and using a dipole-like approximation.
Next, we add the dynamics of the point defects.
The structure of the resulting
theory is much like that of an electron interacting with
electro-magnetic fields. An intuitive microscopic
picture of point defects is also
presented which leads to an alternative
derivation of the action descibing a solid with point defects.
Finally, the hydrodynamic equations are derived.
\subsection{Gauge theory of phonons and dislocations}
\label{2A}
We start with deriving a gauge theory that describes
the solid phase at long wavelenghts. To describe point defects we only
have to include dislocations into our theory. Therefore we can ignore
higher gradient elasticity, which would be needed if we also
wanted to describe disclinations. This section is closely related
to previous work done by Kleinert \cite{Kleinert3,Kleinert1} but
differs from it in the following aspects.
First, we do not include higher gradient elasticity.
Second, we consider the more general case of anisotropic solids and
third, we explicitly remove the unphysical gauge
degrees of freedom in the resulting theory of `quantum defect dynamics'.

The Euclidian action for a solid with dislocations of
arbitrary crystaline symmetry is given by
\cite{Kleinert3,Kleinert1}

\begin{eqnarray}
\label{action}
S[u_{i}] & = &  \int_{0}^{\hbar \beta} d \tau \int d{\bf x} \left \{
\frac{\rho}{2} (\partial_{\tau} u_{i} - \beta_{i})^{2} +
\frac{1}{2} (u_{ij} - \frac{\beta_{ij} + \beta_{ji}}{2})
c_{ijkl} (u_{kl} - \frac{\beta_{kl} + \beta_{lk}}{2}) \right \}
\; ,
\end{eqnarray}
where $u_{ij} = \frac{1}{2}(\partial_{i} u_{j} + \partial_{j} u_{i})$ is the
strain tensor, $c_{ijkl}$ is the elasticity tensor whose structure is
determined by the specific symmetry of the
crystal under consideration and $\rho$ is the average mass density.
This is the most general quadratic action compatibel with the
symmetries of the crystal and the requirement
that the Hamiltonian
of the system transforms under a Gallilean transformation
$({\bf u}, t) \rightarrow ({\bf u + v}t, t)$ as
$H \rightarrow H + {\bf p} \cdot {\bf v} + M {\bf v}^{2}$,
with ${\bf p}$ the total momentum of the
crystal and M it's total mass.
The latter determines the form of the kinetic energy.

The dislocations mentioned above
are topological defects, which exist because the
displacement field is multivalued. In much the same way, vortices in a
superfluid are a consequence of the multivaluedness
of the phase field.\cite{Kleinert3}
The multivaluedness of a displacement field describing a dislocation becomes
apparent when writing down what can be seen as the definition of a
dislocation, namely

\begin{equation}
\oint_{C} d u_{i} = b_{i} \; .
\end{equation}
Here, $C$ is a contour enclosing a dislocation line ${\cal L}$
and ${\bf b}$ is the so-called
Burgers vector measuring the strength of the dislocation.
This equation can be written in a differential form as follows

\begin{equation}
\begin{array}{lcl}
\varepsilon_{ijk} \partial_{j} \partial_{k} u_{l} & = & b_{i}
\delta_{l}({\bf x}, {\cal L}) \equiv \alpha_{il} \; ,
\end{array}
\end{equation}
where $\alpha_{il}$ is called the dislocation density.
It is analogous to
Amp\'ere's law $\varepsilon_{ijk} \partial_{j} B_{k} = J_{i}$ and
states that the displacement field is non-integrable along the line
${\cal L}$ because here
the dislocation gives a delta function contribution.
If the line ${\cal L}$
is parametrized by ${\bf x}(s)$, the delta function
along ${\cal L}$ is defined by

\begin{equation}
\delta_{i}({\bf x}, {\cal L}) = \int ds
\frac{\partial x_{i}(s)}{\partial s} \delta({\bf x} - {\bf x}(s)) \; .
\end{equation}

If we would use in our calculations these multivalued displacement fields,
the action would be given by Eq. (\ref{action}) with
$\beta_{i} = \beta_{ij} = 0$. However, to perform a path integral
over the $u_{i}$ it is much more convenient to use
a singlevalued displacement field which takes values on the real axis.
The unphysical singular
contributions to the derivatives of a singlevalued displacement field
which describes dislocations in a solid are compensated by
substracting the quantities $\beta_{ij}$ and $\beta_{i}$.
The relation between these quantities and the dislocation line ${\cal L}$ is
conveniently visualized by the Volterra construction, which we
now briefly explain.
Given a solid without imperfections, a dislocation can be created by
removing from this
solid a volume ${\cal V}$ and drawing the boundary of the volume together,
thus forming a surface ${\cal S}$ with boundary ${\cal L}$,
and restoring the crystaline symmetry
everywhere except at this boundary ${\cal L}$.

The singlevalued displacement field which describes a dislocation created by
this construction is discontinuous across the surface ${ \cal S}$ with a
jump in the displacement field that is equal to the Burgers vector ${\bf b}$.
This discontinuity gives a delta function contribution to the gradient
of the displacement field which is called the plastic distortion
and is given by

\begin{eqnarray}
\beta_{ij} & = & b_{j} \delta_{i}({\bf x},{\cal S}) \nonumber \\
& = & b_{j} \int_{\cal S} dS_{i} \; \delta({\bf x} - {\bf x}(u,v)) \; .
\end{eqnarray}
The integral measure is defined by
$dS_{i} = \varepsilon_{ijk} \partial_{u} x_{j} \partial_{v} x_{k} du dv$ if
the surface ${\cal S}$ is parametrized by ${\bf x}(u,v)$.
Furthermore, if the dislocation line ${\cal L}$ is moving with a speed
${\bf v}$, the time derivative of the displacement field
gives a delta function contribution of $\beta_{i} = v_{j} \beta_{ji}$.
If we are not on a dislocation line,
the physical values of the spatial and time derivatives of the displacement
field should be continuous and are therefore given by

\begin{eqnarray}
(\partial_{i} u_{j})^{phys} & = & \partial_{i} u_{j} - \beta_{ij} \nonumber \\
(\partial_{\tau} u_{j})^{phys} & = & \partial_{\tau} u_{j} -
\beta_{j} \; .
\end{eqnarray}
The value of these physical quantities equals what one would get by using the
multivalued version of the displacement field to calculate the spatial and
time derivatives.\cite{Zippelius}
We thus see that the action introduced at the beginning of this section
is indeed just the classical action for a perfect crystal
straightforwardly generalized to include dislocations.
\cite{Hirth,Landau}

To be able to actually calculate the interaction
between the phonons and the defects we write the action in a
canonical form. We do this by introducing two new
fields by means of a Hubbard-Stratonovich
transformation.\cite{Negele} Physically these field are
the stress tensor
$\sigma_{ij}$ and momentum density $p_{i}$ which are canonical to
$u_{ij}^{phys} \equiv (\partial_{i} u_{j})^{phys}
- (\partial_{j} u_{i})^{phys}$ and
$(\partial_{\tau} u_{i})^{phys}$ respectively.
It amounts to adding to the Lagrangian density in Eq.
(\ref{action}) the quadratic terms

\begin{eqnarray}
\nonumber
\frac{1}{2 \rho}
\left( p_{i} - i \rho(\partial_{\tau} u_{i})^{phys}
\right)^{2}
\end{eqnarray}
and

\begin{eqnarray}
\nonumber
\frac{1}{2}
\left(
\sigma_{ij} + i u_{gh}^{phys} c_{ghij}
\right)
c^{-1}_{ijkl}
\left(
\sigma_{kl} + ic_{klmn} u_{mn}^{phys} \right) \; .
\end{eqnarray}
Furthermore, to obtain the path integral representation of the
partition function ${\cal Z}$ we have to add
functional integrals over the momentum density
$p_{i}$ and the stress tensor $\sigma_{ij}$.
Note that the action containes only the symmetric part of the stress tensor,
and we should therefore only perform the path integral over the symmetric part
of $\sigma_{ij}$. Note also that $c_{ijkl}^{-1}$ is symmetric under
the exchanges $i \leftrightarrow j$ and $k \leftrightarrow l$,
and  is defined by

\begin{equation}
\label{definition inverse}
c_{ijkl} c^{-1}_{klmn} \equiv \frac{1}{2} (\delta_{im} \delta_{jn} +
\delta_{in} \delta_{jm}) \; .
\end{equation}
The action we find after
these transformations reads

\begin{eqnarray}
\label{action 100}
S[p_{i},\sigma_{ij},u_{i}] =
\int_{0}^{\hbar \beta} d \tau \int d{\bf x}
\left\{ \frac{p_{i}^{2}}{2 \rho}
+ \sigma_{ij} c^{-1}_{ijkl} \sigma_{kl} - ip_{i} (\partial_{\tau}
u_{i})^{phys} + i\sigma_{ij} u_{ij}^{phys} \right\} \; .
\end{eqnarray}

We now integrate out the displacement field,
which leads to the  constraints

\begin{equation}
\label{constraints}
\partial_{\tau} p_{j} = \partial_{i} \sigma_{ij} \; .
\end{equation}
This is Newton's law.
In order to automatically satisfy these constraints we rewrite the fields
$p_{i}$ and
$\sigma_{ij}$ in terms of new fields $A_{ij}$ and $F_{ij}$ by

\begin{eqnarray}
\label{gauge.field.}
\sigma_{ij} & = & \partial_{\tau} F_{ij} + \varepsilon_{ikl}
\partial_{k} A_{lj} \nonumber \\  p_{j} & = & \partial_{i} F_{ij} \; .
\end{eqnarray}
Substituting these in the interaction, i.e. the last two terms in
the right hand side of Eq.
(\ref{action 100}), and performing some partial integrations, we find
that this part of the action can be
written in terms of the dislocation density $\alpha_{ij}$ and
dislocation current density
$J_{mlj} \equiv v_{m} \alpha_{lj} $, as

\begin{equation}
S_{int}[A_{ij},F_{ij}] = \int_{0}^{ \hbar \beta}
d\tau \int d{\bf x} \left \{ -i
A_{ij} \alpha_{ij} - i F_{ij} \varepsilon_{iml} J_{mlj} \right \} \; .
\end{equation}

In the process of rewriting the action we have ended up with to many degrees of
freedom. These unphysical degrees of freedom manifest themselves in
the fact that the new fields $A_{ij}$ and $F_{ij}$ are
gauge fields. Indeed, the expressions for
$\sigma_{ij}$ and $p_{i}$ are invariant
under the gauge transformations

\begin{eqnarray}
\label{gauge transformations}
F_{ij} & \rightarrow & F_{ij} + \varepsilon_{ikl} \partial_{k} \Lambda_{lj}
\nonumber \\
A_{lj} & \rightarrow & A_{lj} + (\partial_{\tau} \Lambda_{lj} - \partial_{l}
\Lambda_{\tau j}) \; .
\end{eqnarray}
At first sight one might therefore think that the gauge freedom
removes $12$ degrees of
freedom. However, we note that these gauge transformations are
themselves invariant under a gauge transformation, which reduces the
number of gauge degrees of freedom.
Indeed, the gauge transformations are invariant under

\begin{eqnarray}
\label{double.gauge.freedom.}
\Lambda_{lj} & \rightarrow &
\Lambda_{lj} + \partial_{l} \lambda_{j} \nonumber \\
\Lambda_{\tau j} & \rightarrow &
\Lambda_{\tau j} + \partial_{\tau} \lambda_{j}
\; .
\end{eqnarray}
As a result the gauge freedom in Eq. (\ref{gauge transformations})
only removes $12-3 = 9$ degrees of freedom
in the expressions for $p_{i}$ and $\sigma_{ij}$.
Therefore we are left with $18-9 = 9$ degrees of freedom in the
fields $A_{ij}$ and $F_{ij}$, which is exactly what we expect because there
are $3$ degrees of freedom present in $p_{i}$, $9$ in $\sigma_{ij}$ and the
constraints in Eq. (\ref{constraints}) remove $3$ of these.
Note that we should also demand $\sigma_{ij}$ to be symmetric, which will
remove another $3$ degrees of freedom.
This means that we end up with $6$ physical
degrees of freedom, corresponding to the usual $6$ phonon modes.

In order to extract physically relevant information we will have to
remove the gauge-degrees of freedom, i.e. fix the gauge.
Before we embark on this problem however, we will prove the following
equalities which we will need later on
when deriving the hydrodynamic equations of motion for
a solid with point defects. They are

\begin{eqnarray}
\label{expectation.values}
\langle u_{ij}^{phys} \rangle & = & i c^{-1}_{ijkl}
\langle \sigma_{kl} \rangle
\nonumber \\
\langle (\partial_{\tau} u_{i})^{phys} \rangle & = &
- \frac{i \langle p_{i} \rangle}{\rho} \; .
\end{eqnarray}
The proof is given by adding to the action in Eq. (\ref{action})
source terms proportional to the currents $K_{ij}$ and $K_{i}$

\begin{eqnarray}
\nonumber
\int_{0}^{\hbar \beta} d\tau \int d{\bf x}
\left\{
K_{ij} u_{ij}^{phys} + K_{i} (\partial_{\tau} u_{i})^{phys}
\right\} \; .
\end{eqnarray}
Expectation values of
$f ( u_{ij}^{phys},
(\partial_{\tau} u_{i})^{phys} )$
are now easily calculated as follows

\begin{equation}
\langle f ( u_{ij}^{phys},
(\partial_{\tau} u_{i})^{phys} ) \rangle =
f(\frac{\partial}{\partial{K_{ij}}}, \frac{\partial}{\partial{K_{i}}})
\ln {\cal Z}(K_{ij},K_{i}) |_{K_{ij} = K_{i} = 0} \; ,
\end{equation}
where ${\cal Z}(K_{ij}, K_{i})$ denotes the partition function with
non-vanishing source terms.
Again performing a Hubbard-Stratonovich transformation we get

\begin{eqnarray}
S[u_{i}] = \int_{0}^{\hbar \beta} d \tau \int d{\bf x} \left\{
\frac{p_{i}^{2}}{2 \rho} + \frac{1}{2} \sigma_{ij} c_{ijkl}^{-1} \sigma_{kl}
- i p_{i} \left( (\partial_{\tau} u_{i})^{phys} +
\frac{K_{i}}{\rho} \right)
+ i\sigma_{ij} (u_{ij}^{phys} + c_{ijkl}^{-1} K_{kl})
\right. \nonumber \\ \left.
- \frac{K_{i}^{2}}{2 \rho} - K_{ij} c_{ijkl}^{-1} K_{kl} \right\} \; .
\end{eqnarray}
Eq. (\ref{expectation.values}) now follow by differentiation.

After this digression, we return to
the elimination of the non-physical degrees of freedom
present in the action $S = S_{0} + S_{int}$ and reexamine Eq.
(\ref{gauge transformations}).
As mentioned, this gauge transformation is invariant under
the transformations in Eq. (\ref{double.gauge.freedom.}).
We use the latter invariance to choose a gauge
in which $\Lambda_{\tau j} = 0$, which is always possible by
letting $\lambda_{j}$ satisfy

\begin{equation}
\lambda_{j} = - \int_{0}^{\tau} \Lambda_{\tau j}(\tau') d \tau' \; .
\end{equation}
In this gauge our original gauge transformation reduces to

\begin{eqnarray}
F_{ij} & \rightarrow & F_{ij} + \varepsilon_{ikl} \partial_{k} \Lambda_{lj}
\nonumber \\
A_{lj} & \rightarrow & A_{lj} - \partial_{\tau} \Lambda_{lj} \; .
\end{eqnarray}
In order for $\sigma_{ij}$ to be symmetric,
we use part of this residual gauge freedom to choose $F_{ij}$ symmetric.
In addition, we introduce the fields $\chi_{ij}$ by means of

\begin{equation}
A_{lj} = \varepsilon_{jmn} \partial_{m} \chi_{ln} \; .
\end{equation}
If we now take $\chi_{ln}$ to be symmetric, $\sigma_{ij}$ will
also be symmetric.
In terms of these fields the free action $S_{0}$ becomes

\begin{eqnarray}
S_{0}[F_{ij},\chi_{ij}] & = & \int_{0}^{\hbar \beta}
d\tau \int d{\bf x} \left\{
\frac{(\partial_{i} F_{ij})^{2}}{2 \rho} \right. \nonumber \\
& & \left. + \frac{1}{2}(\partial_{\tau} F_{ij}
+ \varepsilon_{ikl} \varepsilon_{jmn} \partial_{k} \partial_{m} \chi_{ln})
c_{ijkl}^{-1}
(\partial_{\tau} F_{kl} + \varepsilon_{kpq} \varepsilon_{jrs} \partial_{p}
\partial_{r} \chi_{qs} ) \right \} \; .
\end{eqnarray}
At this point $F_{ij}$ and $\chi_{ij}$ both contain
$6$ degrees of freedom.
We are thus left with $12 - 6  = 6$ non-physical
degrees of freedom which somehow correspond to 6
degrees of freedom in $\Lambda_{ij}$.

To eliminate the remaining unphysical degrees of freedom,
we expand the Fourier transform of the
fields $F_{ij}$ and $\chi_{ij}$ in the helicity-basis
$\{ e_{ij}^{(s,h)} \}$.\cite{Kleinert3}
If we take a direct product of momentum space
${\cal P}$ with itself, i.e. ${\cal P} \otimes {\cal P}$,
the helicity basis
is defined as the irreducible representations of the rotation group in this
space. From group theory we know that they form a
complete set.\cite{Hamermesh}
Hence we can develop a given tensor field in this basis leading to

\begin{eqnarray}
\chi_{ij}({\bf k}) & = & \sum_{s,h} e^{(s,h)}_{ij}(\hat {\bf k})
\chi^{(s,h)}({\bf k}) \nonumber \\
F_{ij}({\bf k}) & = & \sum_{s,h} e^{(s,h)}_{ij}(\hat {\bf k})
F^{(s,h)}({\bf k}) \; .
\end{eqnarray}
Because of the symmetry of $\chi_{ij}$ and $F_{ij}$ the 6 non-zero
components are $(s,h)=\{ (0,0), (2,0), (2, \pm 1), (2,\pm 2) \}$.
To identify the surviving physical helicity components,
we note that the expression $\varepsilon_{ikl}
\varepsilon_{jmn} \partial_{k}
\partial_{m} \chi_{ln}$ is symmetric, traceless and
invariant under the following transformation

\begin{equation}
\label{helicity.gauge.}
\chi_{ln} \rightarrow \chi_{ln} +
\partial_{l} \xi_{n} + \partial_{n} \xi_{l} \; .
\end{equation}
If we choose as a basis in Fourier space
$\{ \hat{\bf k}_{n}, e^{(1,1)}_{n}(\hat{\bf k}),
e^{(1,-1)}_{n}(\hat {\bf k}) \}$ and develop $\xi_{n}$
in terms of this basis, this transformation reads up to a factor $k$

\begin{eqnarray}
{\hat k}_{l} \xi_{n} + {\hat k}_{n} \xi_{l} & = &
2 {\hat k}_{l} \hat k_{n} \xi^{(0)} +
( {\hat k}_{l} e^{(1,1)}_{n} + {\hat k}_{n} e^{(1,-1)}_{l}) \xi^{(1)} +
( {\hat k}_{l} e^{(1,-1)}_{n} + {\hat k}_{n} e_{l}^{(1,-1)}) \xi^{(-1)}
\nonumber \\
& = & \frac{2}{\sqrt{3}}(\sqrt{2} e^{(2,0)}_{ln} +
e^{(0,0)}_{ln}) \xi^{(0)} +
\sqrt{2} e^{(2,1)}_{ln} \xi^{(1)} + \sqrt{2} e_{ln}^{(2,-1)} \xi^{(2)} \; .
\end{eqnarray}
{}From this expression we see that if we choose a new basis
in which to develop $\chi_{ln}$ given by

\begin{equation}
\label{new basis}
\{ e_{ln}^{(2,2)}, e_{ln}^{(2,-2)}, e_{ln}^{(2,1)}, e_{ln}^{(2,-1)},
e_{ln}^{L}, e_{ln}^{L'} \} \; ,
\end{equation}
where

\begin{eqnarray}
e_{ln}^{L} & = & \frac{1}{\sqrt{3}}(-e_{ln}^{(2,0)} + \sqrt{2} e_{ln}^{(0,0)})
= \frac{1}{\sqrt{2}}(\delta_{ln} - \hat k_{l} \hat k_{n}) \nonumber \\
e_{ln}^{L'} & = & \frac{1}{\sqrt{3}}(\sqrt{2} e_{ln}^{(2,0)} +
e_{ln}^{(0,0)}) =
\hat k_{l} \hat k_{n} \; ,
\end{eqnarray}
the components of $\chi_{ln}$ corresponding to $\{(2,1), (2,-1), L' \}$
are unphysical and disappear from the action
because they correspond to a gauge transformation.
The coordinate transformation from the old to the new basis
is unitary and hence the new basis is also orthonormal and
complete in the space of symmetric second rank tensors.
In addition the elements $\{ (2,2),(2,-2),L \}$ satisfy

\begin{equation}
k_{l} e_{ln}^{(L)} = k_{l} e_{ln}^{(2,2)} = k_{l} e_{ln}^{(2,-2)} = 0 \; .
\end{equation}
This means there are only 3 dynamical degrees of freedom left in
$\partial_{i} F_{ij}$ corresponding to the helicity components
$\{ (2,1),(2,-1),L' \}$.

We now Fourier transform the action and
expand the fields $F_{ij}$ and $\chi_{ij}$ in terms of the basis
in Eq. (\ref{new basis}). We get

\begin{eqnarray}
S_{0}[F_{ij},\chi_{ij}] = \int_{0}^{\hbar \beta}
d\tau \int \frac{d{\bf k}}{(2 \pi)^{3}} \left\{
\frac{1}{2 \rho} \left| ik_{i}(e_{ij}^{(2,1)} F^{(2,1)} +
e_{ij}^{(2,-1)}F^{(2,-1)} + e_{ij}^{L'}F^{L'}) \right| ^{2} \right.
\nonumber \\
\left. + \frac{1}{2} \left[
\sum_{s,h} e_{ij}^{(s,h)} \partial_{\tau} F^{(s,h)} +
k^{2}(e_{ij}^{(2,2)} \chi^{(2,2)} +
e_{ij}^{(2,-2)} \chi^{(2,-2)} - e_{ij}^{L} \chi^{L})\right]^{*} c_{ijkl}^{-1}
[ij \leftrightarrow kl] \right\} \; ,
\end{eqnarray}
where $[ij \leftrightarrow kl]$ denotes the part between brackets with
the indicated interchange of indices.
Up to now we have removed all but three unphysical degrees of freedom.
By introducing $\chi_{ij}$ and choosing a gauge in which $F_{ij}$
and $\chi_{ij}$ are both symmetric, the stress tensor was made symmetric.
Hence we were left with $12$ degrees of freedom.
Then we identified $3$ unphysical gauge
degrees of freedom in $\chi_{ij}$ corresponding to the helicity components
$\{(2,1), (2,-1), L' \}$, which reduced the remaining number
degrees of freedom to $9$. Hence we expect a residual gauge freedom to be
present in the above action corresponding to three unphysical degrees of
freedom. As is apparent in the expression for $S_{0}$, this
is indeed the case and the remaining gauge freedom corresponds to

\begin{eqnarray}
F^{(s,h)} & \rightarrow & F^{(s,h)} + k^{2} \Lambda^{(s,h)} \nonumber \\
\chi^{(s,h)} & \rightarrow & \chi^{(s,h)} - \partial_{\tau}
\Lambda^{(s,h)} \; ,
\end{eqnarray}
for $(s,h) = \{ (2,2), (2,-2), L \}$. We will see below
that this gauge freedom is also present in $S_{int}$.
To remove this remaining freedom we introduce three new, invariant, fields

\begin{eqnarray}
\label{new.fields}
\chi'^{(2,2)} & = & \chi^{(2,2)} +
\frac{\partial_{\tau} F^{(2,2)}}{k^{2}} \nonumber \\
\chi'^{(2,-2)} & = & \chi^{(2,-2)} + \frac{\partial_{\tau}
F^{(2,-2)}}{k^{2}} \\
\chi'^{L} & = & \chi^{L} - \frac{\partial_{\tau} F^{L}}{k^{2}} \nonumber \; .
\end{eqnarray}
If we furthermore realize that

\begin{eqnarray}
k_{i} e^{(2,1)}_{ij} & = & \frac{k}{\sqrt{2}}e_{j}^{(1,1)} \nonumber \\
k_{i} e^{(2,-1)}_{ij} & = & \frac{k}{\sqrt{2}}e_{j}^{(1,-1)} \\
k_{i} e^{L'}_{ij} & = & k \hat k_{j} \nonumber \; ,
\end{eqnarray}
we obtain the
final expression for $S_{0}$ which contains precisely 6 dynamical
degrees of freedom.

Before explicitly writing down $S_{0}$
we introduce a new compact notation which also
simplifies the algebraic manipulations
involved in the remainder of this article.
We define the following quantities

\begin{eqnarray}
  \vec{F} =
  \left(
  \begin{array}{c}
    F^{(2,1)} \\
    F^{(2,-1)} \\
    F^{L'} \\
  \end{array}
  \right) \;  ;  \;
  \vec{\chi'} =
  \left(
  \begin{array}{c}
    \chi'^{(2,2)} \\
    \chi'^{(2,-2)} \\
    -\chi'^{L} \\
  \end{array}
  \right) \; ; \;
  \vec{e}_{ij}^{(1)} =
  \left(
  \begin{array}{c}
    e^{(2,1)}_{ij} \\
    e^{(2,-1)}_{ij} \\
    e^{L'}_{ij} \\
  \end{array}
  \right) \;  ;  \;
  \vec{e}_{ij}^{(2)} =
  \left(
  \begin{array}{c}
    e^{(2,2)}_{ij} \\
    e^{(2,-2)}_{ij} \\
    e^{L}_{ij} \\
  \end{array}
  \right)
\end{eqnarray}
and

\begin{eqnarray}
\label{C}
\begin{array}{lcl}
  A_{\mu \nu} & = & ( \vec{e}_{ij}^{(1)} )_{\mu} c_{ijkl}^{-1}
  ( \vec{e}_{kl}^{(1)} )_{\nu} \\
  B_{\mu \nu} & = & ( \vec{e}_{ij}^{(1)} )_{\mu} c_{ijkl}^{-1}
  ( \vec{e}_{kl}^{(2)} )_{\nu} \\
  C_{\mu \nu} & = & ( \vec{e}_{ij}^{(2)} )_{\mu} c_{ijkl}^{-1}
  ( \vec{e}_{kl}^{(2)} )_{\nu} \\
\end{array}
\; \; \; ; \; \; \;
A' & = &
  \left(
  \begin{array}{ccc}
    1 & 0 & 0 \\
    0 & 1 & 0 \\
    0 & 0 & 2
  \end{array}
  \right) \; .
\end{eqnarray}
Indices refering to the abstract vector space introduced
above, are denoted by Greek
symbols to distinguish them from their real space counterparts.
This allows us to write $S_{0}$ in the following way

\begin{eqnarray}
\label{new.notation.action}
S_{0}[\vec{F}, \vec{\chi}'] =
\frac{1}{2}
\int_{0}^{\hbar \beta} d\tau \int
\frac{d{\bf k}}{(2 \pi)^{3}}
\left(
\begin{array}{c}
\vec{F} \\
\vec{\chi'}
\end{array}
\right)^{*} \cdot
\left(
\begin{array}{cc}
\frac{A' k^{2}}{2 \rho} -
A \partial_{\tau}^{2} & - k^{2} B^{\dagger} \partial_{\tau} \\
k^{2} B \partial_{\tau} & k^{4} C
\end{array}
\right) \cdot
\left(
\begin{array}{c}
\vec{F} \\
\vec{\chi'}
\end{array}
\right) \; ,
\end{eqnarray}
Note the minus signs, which arise from partial integration.

Next, we also have to write the interaction in terms
of the physical fields $\vec{F}$ and $\vec{\chi}'$.
The interaction in terms of
$F_{ij}$ and $\chi_{ij}$ is given by

\begin{equation}
\label{interaction0}
S_{int}[\chi_{ij},F_{ij}] =
\int d\tau \int d{\bf x}
\left\{
-i \varepsilon_{jmn} \partial_{m} \chi_{in} \alpha_{ij} -
i F_{ij} \varepsilon_{iml} J_{mlj}
\right\} \; .
\end{equation}
It is not immediately obvious from this equation that the interaction can
be rewritten in terms of $\vec{F}$ and $\vec{\chi'}$.
Therefore we will show this explicitly.
After partially integrating the first part of Eq.
(\ref{interaction0}), the field $\chi_{ij}$ interacts with
$\varepsilon_{jmn} \partial_{m} \alpha_{ij} =
\varepsilon_{jmn} \partial_{m} \varepsilon_{ikl} \partial_{k} \beta_{lj}$
which is symmetric and traceless and therefore
has only the helicity components $\{(2,2),(2,-2),L'\}$.
Next we rewrite the second term term $ -i F_{ij}
\varepsilon_{iml} J_{mlj}$ as follows

\begin{eqnarray}
\label{rewritten interaction}
-i F_{ij} \varepsilon_{iml} J_{mlj} =
 -i F_{ij}
\left[
\left(
\delta_{il} - \frac{\partial_{i} \partial_{l}}{\partial^{2}}
\right)
\left(
\delta_{kj} - \frac{\partial_{k} \partial_{j}}{\partial^{2}}
\right) \varepsilon_{lmn} J_{mnk} + \right. \nonumber \\ \left.
\frac{\partial_{i} \partial_{l}}{\partial^{2}}
\varepsilon_{lmn} J_{mnj} +
\left(
\delta_{il} - \frac{\partial_{i} \partial_{l}}{\partial^{2}}
\right)
\frac{\partial_{k} \partial_{j}}{\partial^{2}}
\varepsilon_{lmn} J_{mnk}
\right]  \; .
\end{eqnarray}
The second and third term on the right hand side of the above equation
will give an interaction
with $F^{(2,1)}$, $F^{(2,-1)}$ and $F^{(L')}$ as is obvious from
the fact that the
contractions $\partial_{i} F_{ij}$ and $\partial_{j} F_{ij}$
annihilate the components $\{(2,2), (2,-2), L \}$.
The first term together with the interaction term
involving $\chi_{ij}$ will reduce to an interaction with the new fields
introduced in Eq. (\ref{new.fields}), i.e.
$\vec{\chi'}$. To see this we use

\begin{equation}
\partial^{2} \delta_{ij} - \partial_{i} \partial_{j} =
\varepsilon_{iml} \partial_{m} \varepsilon_{lkj} \partial_{k} \; ,
\end{equation}
and substite this in the first term on the right-hand side of
Eq. (\ref{rewritten interaction}). We obtain

\begin{eqnarray}
& & -i \int d{\bf x}
\frac{F_{ij}}{\partial^{4}}
(\varepsilon_{ipq} \partial_{p} \varepsilon_{qrl} \partial_{r})
(\varepsilon_{kst} \partial_{s} \varepsilon_{tvj} \partial_{v})
\varepsilon_{lmn} J_{mnk} \nonumber \\
& = & i \int d{\bf x}
\left( \varepsilon_{qpi} \partial_{p} \varepsilon_{tvj} \partial_{v}
\frac{F_{ij}}{\partial^{4}} \right)
\varepsilon_{kst} \partial_{s} \varepsilon_{qrl} \partial_{r}
\varepsilon_{lmn} J_{mnk} \nonumber \\
& = & -i \int d{\bf x}
\left( \varepsilon_{qpi} \partial_{p} \varepsilon_{tvj} \partial_{v}
\frac{F_{ij}}{\partial^{4}} \right)
\varepsilon_{kst} \partial_{s} \partial_{\tau} \alpha_{qk} \nonumber \\
& = & i \int d{\bf x}
\left( \varepsilon_{qpi} \partial_{p} \varepsilon_{tvj} \partial_{v}
\frac{\partial_{\tau} F_{ij}}{\partial^{4}} \right)
\varepsilon_{kst} \partial_{s} \alpha_{qk} \; ,
\end{eqnarray}
where we have used

\begin{eqnarray}
\varepsilon_{qrl} \partial_{r} \varepsilon_{lmn} J_{mkl} & = &
\partial_{r}
(\delta_{qm} \delta_{rn} - \delta_{qn} \delta_{rm})
v_{m} \alpha_{nk} \nonumber \\
& = & \partial_{\tau} \alpha_{qk} \; .
\end{eqnarray}
We see that there is indeed only an interaction with the
$(2,2)$, $(2,-2)$ and $L$ components of $F_{ij}$.
Together with the expansion for $-i \varepsilon_{jmn} \partial_{m}
\chi_{in} \alpha_{ij}$ these precisely form the fields $\chi'^{(s,h)}$.
Inserting all this into $S_{int}$,
the interaction between the dislocations and the physical
fields describing the phonon modes finally becomes

\begin{eqnarray}
\label{interaction}
S_{int}[\vec{F},\vec{\chi}']
& = & i \int d\tau \int \frac{d {\bf k}}{(2 \pi)^{3}} \left\{
(e_{ln}^{(2,2)} \chi'^{(2,2)} +
e_{ln}^{(2,-2)} \chi'^{(2,-2)} + e_{ln}^{L} \chi'^{L})\varepsilon_{jmn}
[i k_{m} \alpha_{lj}]^{*} - \right. \nonumber \\
& & (e_{ij}^{(2,1)} F^{(2,1)} + e_{ij}^{(2,-1)} F^{(2,-1)} -
e_{ij}^{(L')} F^{(L')} ) \hat k_{i} \hat k_{l}
\varepsilon_{lmn} J_{mnj}^{*} +  \nonumber \\
& & \left. (e_{ij}^{(2,1)} F^{(2,1)} + e_{ij}^{(2,-1)} F^{(2,-1)} -
e_{ij}^{(L')} F^{(L')} ) \varepsilon_{ipq}
\hat k_{p} \hat k_{k} \hat k_{j} \hat k_{m} J_{mqk}^{*}
\right\} \; .
\end{eqnarray}

{}From this interaction between phonons and a dislocation loop
we now want to derive the interaction between phonons and a point defect.
To do so, we consider a point defect to be a
dislocation loop that shrinks to zero radius.
The Volterra construction shows that one creates a dislocation
by removing a volume ${\cal V}$ from the crystal.
When shrinking the dislocation loop,
this volume finally ends up being the volume of a single atom. In this way
one can thus remove or add the volume of a single atom,
which results in creating a vacancy or an interstitial.
Alternatively, one can take the long wavelength limit, in
which the dislocation will effectively look like a point defect.
However, naively applying the above procedure yields zero interaction
between the point defects and the phonons because a point defect has no
Burgers vector.
Indeed, assuming that the wavelength of the fluctuations of the
phonon fields are much larger than the radius of a dislocation allows for
the action to be coarse grained. The physical fields $\vec{F}$ and
$\vec{\chi}$ then interact with something proportional to

\begin{eqnarray}
\nonumber
\int_{{\cal V}} d{\bf x} \; \alpha_{ij} \; .
\end{eqnarray}
However, the above quantity is zero and therefore there is
no interaction with the phonon fields in a first approximation.
We thus conclude that in order to find an interaction we need to have
a gradient of the fields $\vec{F}$ and $\vec{\chi}$ over the radius of a
dislocation loop. This is
analogous to the well known multipole expansion in electrodynamics,
where a dipole has no netto charge but interacts with the gradient of the
electro-magnetic potentials.
Assuming the physical fields to be slowly varying over the region of
nonzero dislocation density we can perform a gradient expansion
that leads to an interaction of the following form

\begin{equation}
S_{int} = i \int_{0}^{\hbar \beta} d\tau \int \frac{d {\bf k}}{(2 \pi)^{2}}
\left\{
k^{2} \vec{a} \cdot \vec{\chi'} N_{\Delta}^{*} +
k \vec{F} \cdot M \cdot \vec{J}^{*}
\right\} \; ,
\end{equation}
containing an extra factor of $k$. The defect density is denoted by
$N_{\Delta}$, $\vec{J} = (J^{(0)}, J^{(1)}, J^{(-1)})$ are
the helicity components of the defect current density
${\bf J}$ and $M_{\mu \nu}$ is a matrix.
If we have $n$ defects located at
$\{ {\bf x}^{(n)} \}$,
the defect density $N_{\Delta}$ and defect current density
$J_{i}$ are given by

\begin{eqnarray}
N_{\Delta}({\bf x}) & = & \sum_{n} q^{(n)} \delta({\bf x} - {\bf x}^{(n)})
\nonumber \\
J_{i}({\bf x}) & = & \sum_{n} i \partial_{\tau}{\bf x}^{(n)} q^{(n)}
\delta({\bf x} - {\bf x}^{(n)}) \; .
\end{eqnarray}
In general there can be both vacancies and interstitials present,
and the charge $q$ distinguishes between vacancies ($q = -1$)
and interstitials ($q = 1$).
Thus, the netto defect density $N_{\Delta}$ is in fact
the difference between the interstitial density and the vacancy density
$N_{\Delta}^{int} - N_{\Delta}^{def}$, which is
conserved because defects and interstitials can locally only be
created in pairs.
The defect density and defect current density
therefore satisfy a continuity equation

\begin{equation}
\partial_{\tau} N_{\Delta} = i \partial_{i} J_{i} \; .
\end{equation}

Up to this point, no specific crystaline symmetry has been assumed.
However, the explicit form of the interaction is determined by the symmetry
of the crystal under consideration. This symmetry constrains
the coefficients $\vec{a}$ and the matrix $M$. In principle it should be
possible to expicitly take the limit of a dislocation loop shrinking to
zero in the interaction given by Eq. (\ref{interaction}). The symmetry
of the crystal is then contained in $\alpha_{ij}$ and $J_{ijk}$, because the
Burgers vector can only be a lattice vector.
There are however subtilities involved in doing this,
and the vector $\vec{a}$ and matrix $M$ will therefore be
determined from symmetry considerations. Because we are especially
interested in the behavior of solid $^{4}$He, we consider from
this point on the special case of a hexagonally close
packed (HCP) crystal structure.
The associated symmetry group is ${\cal C}_{6h}$, which contains rotations
about the $c$-axis and reflections in the $ab$-plane. The elasticity tensor
$c_{ijkl}$ for this symmetry containes $5$ constants, the analogues
of the Lam\'e constants $\lambda$ and $\mu$ in the isotropic case.

In what follows we need the field equations for $\vec{F}$ and $\vec{\chi}'$
which follow from the complete action $S_{0} + S_{int}$ and are given by
\begin{eqnarray}
\label{field equations}
\partial_{\tau}^{2} \vec{F} & = & \frac{k^{2}}{2 \rho} P \cdot
\left( A' \cdot \vec{F} + i \frac{2 \rho}{k} M \cdot \vec{J} \right) -
i Q^{\dagger} \cdot \vec{a} \partial_{\tau} N_{\Delta} \nonumber \\
\partial_{\tau} \vec{\chi}' & = & \frac{1}{2 \rho} Q \cdot
\left( A' \cdot \vec{F} + i \frac{2 \rho}{k} M \cdot \vec{J} \right) -
\frac{i}{k^{2}} R \cdot \vec{a} \partial_{\tau}
N_{\Delta} \; ,
\end{eqnarray}
where

\begin{equation}
\left(
\begin{array}{cc}
P & Q^{\dagger} \\
Q & R
\end{array}
\right) \cdot
\left(
\begin{array}{cc}
A & B^{\dagger} \\
B & C
\end{array}
\right)
=
\left(
\begin{array}{cc}
1 & 0 \\
0 & 1
\end{array}
\right) \; .
\end{equation}
To determine the form of the interaction between the phonons and the defects
we calculate the
stress tensor $\sigma_{ij}$ resulting from a single point defect.
We then note that from the symmetry
of the crystal and the fact that a single point-defect has no
orientation it follows that $\sigma_{ij}$ has to be invariant
under the symmetry operations of ${\cal C}_{6h}$, i.e. rotations around the
$c$-axis and reflections in the $ab$-plane.
Denoting a particular symmetry operation by $S$, we get

\begin{eqnarray}
S_{ik} S_{jl} \sigma_{kl}(S^{-1} {\bf x}) & = &
\sigma_{ij}({\bf x})  \; ,
\end{eqnarray}
which in Fourier language reads

\begin{equation}
\label{restriction}
S_{ik} S_{jl} \sigma_{kl}(S^{-1} {\bf k}) = \sigma_{ij}({\bf k})\; .
\end{equation}
In the case of a static defect we can solve for
$\sigma_{ij}$, using the field equations for
$\vec{\chi}'$. These can be derived from Eq.
(\ref{field equations}) and read

\begin{equation}
\label{equation.motion}
\vec{\chi}' = - \frac{i}{k^{2}} C^{-1} \cdot \vec{a}N_{\Delta}
\; .
\end{equation}
With the notation introduced before
$\chi_{\mu}$ and $a_{\nu}$ are vectors, and $C_{\mu \nu}$
is a matrix given by Eq. (\ref{C}).
Inserting this into the equation for $\sigma_{ij}$ with
$\vec{F} = 0$, i.e. considering a static defect, we get

\begin{eqnarray}
\sigma_{ij} & = & k^{2} \vec{e}_{ij}^{(2)} \cdot \vec{\chi}' \nonumber \\
& = & - i \vec{e}_{ij}^{(2)} \cdot C^{-1} \cdot \vec{a}
N_{\Delta} \; .
\end{eqnarray}
This means  that Eq. (\ref{restriction}) translates into

\begin{equation}
\label{restriction2}
i \left\{
[S_{ik} S_{jl} \vec{e}^{(2)}_{kl} (S^{-1} {\bf \hat{k}})]
\cdot [C^{-1}(S^{-1} {\bf \hat{k}})] -
[\vec{e}_{kl}^{(2)}({\bf \hat{k}})]
\cdot [C^{-1}({\bf \hat{k}})]
\right\} \cdot \vec{a} = 0 \; .
\end{equation}

Next we are going to translate this equation into a restriction
on the coefficients $\vec{a}$
in the interaction. In order to do so, we must
choose a particular form for the thus far unspecified
helicity-basis. On each point of the unit sphere we choose
three orthonormal vectors. One in the radial direction, i.e. $\hat{\bf k}$,
the other two in such a way that the vector fields
we get in this way are invariant under rotations about the
c-axis. This cannot be done for
the entire sphere and the points
$(k_{a},k_{b},k_{c})= \{ (0,0,1), (0,0,-1) \}$
are excluded. Therefore the only points where
Eq. (\ref{restriction2}) is not, by construction,
automatically satisfied for rotations about the c-axis,
are indeed these two points.
We only treat $(k_{a},k_{b},k_{c}) = (0,0,1)$,
because the other point does not lead to
any additional restrictions. In this point we choose

\begin{eqnarray}
{\bf e}^{(1)} & = & (1,0,0)  \nonumber \\
{\bf e}^{(2)} & = & (0,1,0) \\
{\bf e}^{(3)} & = & (0,0,1) = \hat{\bf k} \nonumber \; .
\end{eqnarray}
This means that here our helicity-basis becomes

\begin{eqnarray}
 e_{ij}^{L} & = & \frac{1}{\sqrt{2}} \left(
  \begin{array}{ccc}
    1 & 0 & 0 \\
    0 & 1 & 0 \\
    0 & 0 & 0
  \end{array}
 \right)_{ij} \nonumber \\
 e_{ij}^{(2,2)} & = & \frac{1}{2} \left(
  \begin{array}{ccc}
   1 & i & 0 \\
   i & -1 & 0 \\
   0 & 0 & 0
  \end{array}
 \right)_{ij}  \\
 e_{ij}^{(2,-2)} & = & \frac{1}{2} \left(
  \begin{array}{ccc}
   1 & -i & 0 \\
   -i & -1 & 0 \\
   0 & 0 & 0
  \end{array}
 \right)_{ij} \; . \nonumber
\end{eqnarray}
Each of these three matrices transforms according to an
irreducible representation of ${\cal C}_{6h}$.
Thus, according to Schur's lemma,\cite{Hamermesh} they do not mix
under the operator $c_{ijkl}$, which means that $C_{\mu \nu} = 0$ if
$\mu \not= \nu$.
We notice that these matrices transform under rotations over
an angle $\alpha$ about the c-axis (we are only considering the
point (0,0,1)) as

\begin{eqnarray}
 e_{ij}^{L}(\hat {\bf k}) & \rightarrow & e_{ij}^{L}(\hat {\bf k}) \nonumber \\
 e_{ij}^{(2,2)}(\hat {\bf k}) & \rightarrow & e^{i 2 \alpha}
e_{ij}^{(2,2)}(\hat {\bf k}) \\
 e_{ij}^{(2,-2)}(\hat {\bf k}) & \rightarrow & e^{-i 2 \alpha}
 e_{ij}^{(2,-2)}(\hat {\bf k}) \nonumber \; .
\end{eqnarray}
It now follows that

\begin{equation}
C^{-1}_{\mu \nu}(S^{-1} \hat {\bf k}) = C^{-1}_{\mu \nu}(\hat {\bf k}) \; .
\end{equation}
Thus for infinitesimal rotations Eq. (\ref{restriction2}) reduces to

\begin{equation}
\left(
\begin{array}{c}
i 2 \alpha e_{ij}^{(2,2)}\\
-i 2 \alpha e_{ij}^{(2,-2)}\\
0\\
\end{array}
\right) \cdot
\left(
\begin{array}{ccc}
C_{11}^{-1} & 0 & 0 \\
0 & C_{22}^{-1} & 0 \\
0 & 0 & C_{33}^{-1}
\end{array}
\right) \cdot
\left(
\begin{array}{c}
a^{(2,2)}\\
a^{(2,-2)}\\
a^{L}\\
\end{array}
\right) = 0 \; .
\end{equation}
This equation has to be valid for all values of $i$ and $j$.
Therefore we get

\begin{eqnarray}
a^{(2,2)} C_{11}^{-1} - a^{(2,-2)} C_{22}^{-1}  =  0
\; \mbox{for} \; i = j \nonumber \\
a^{(2,2)} C_{11}^{-1} + a^{(2,-2)} C_{22}^{-1}  =  0
\; \mbox{for} \; i \not= j \; .
\end{eqnarray}
The only solution to these equations is

\begin{equation}
a^{(2,2)} = a^{(2,-2)} = 0 \; ,
\end{equation}
which means that our interaction in first instance reduces to

\begin{eqnarray}
\label{point defect action}
S_{int}[\vec{F}, \vec{\chi}'] =
\int_{0}^{\hbar \beta} d\tau \int \frac{d{\bf k}}{(2 \pi)^{3}}
\left\{ k^{2} a^{L} \chi^{L} N_{\Delta}^{*} +
k \vec{F} \cdot M \cdot \vec{J}^{*}
\right\} \; .
\end{eqnarray}

Next we must determine the form of the matrix $M$.
This is done by demanding the following equality to be valid
\begin{equation}
\label{equality two densities}
i \rho \partial_{\tau} \langle (\partial_{i} u_{i})^{phys} \rangle =
i \rho \partial_{i} \langle (\partial_{\tau} u_{i})^{phys} \rangle \; .
\end{equation}
If the displacement field is
singlevalued and continuous everywhere, the left and right hand side of this
equation are two equivalent expressions for $- i \partial_{\tau} \delta \rho$.
Therefore it should be valid when
there are only point defects present. However,
when there are dislocations present Eq. (\ref{equality two densities})
can be shown to be false. It is therefore
not at all obvious that the equality is satisfied at this point, because we
started out with a description including dislocations.
Indeed, the above requirement actually gives a constraint on $M$,
as we will see below.
We first calculate $i \rho \partial_{\tau}\langle
(\partial_{i} u_{i})^{phys} \rangle$ and find

\begin{eqnarray}
\label{rho1}
i \rho \partial_{\tau} \langle (\partial_{i} u_{i})^{phys} \rangle & = & -
\rho \partial_{\tau} c_{iikl}^{-1} \langle \sigma_{kl} \rangle \nonumber \\
& = & - \rho \int \frac{d{\bf k}}{(2 \pi)^{3}}
e^{i\bf{k} \cdot \bf{x}} c^{-1}_{iikl}
\left(
\vec{e}_{kl}^{(1)} \cdot \partial_{\tau}^{2} \langle \vec{F} \rangle +
k^{2} \vec{e}_{kl}^{(2)} \cdot \partial_{\tau} \langle \vec{\chi} \rangle
\right) \nonumber \\
& = &
- \int \frac{d{\bf k}}{(2 \pi)^{3}} e^{i\bf{k} \cdot \bf{x}}
\left\{
k^{2} \langle F^{L'} \rangle - \rho
\left(
\sqrt{2} a^{L} \partial_{\tau} \langle N_{\Delta} \rangle +
k M_{1j} \langle J_{j} \rangle
\right)
\right\} \; ,
\end{eqnarray}
where we used the equations of motion in Eq. (\ref{field equations}),
the completeness relation for the helicity basis in writing

\begin{eqnarray}
\vec{e}_{ij}^{(1)} \cdot Q^{\dagger} + \vec{e}_{ij}^{(2)} \cdot R & = &
\left(
\vec{e}_{ij}^{(1)} \cdot \vec{e}_{kl}^{(1)} + \vec{e}_{ij}^{(2)}
\cdot \vec{e}_{kl}^{(2)} \right)
c_{klmn} \vec{e}^{(2)}_{mn} \nonumber \\
& = & c_{ijmn} \vec{e}^{(2)}_{mn} \nonumber \\
\vec{e}_{ij}^{(1)} \cdot P + \vec{e}_{ij}^{(2)} \cdot Q & = &
\left(
\vec{e}_{ij}^{(1)} \cdot \vec{e}_{kl}^{(1)} + \vec{e}_{ij}^{(2)}
\cdot \vec{e}_{kl}^{(1)} \right)
c_{klmn} \vec{e}^{(2)}_{mn} \nonumber \\
& = & c_{ijmn} \vec{e}^{(1)}_{mn} \; ,
\end{eqnarray}
and the fact that only $e^{L}_{ij}$ and $e^{L'}_{ij}$ are traceless.
If we compare the result in Eq. (\ref{rho1}) with
the expression for  $i \rho \partial_{i}
\langle (\partial_{\tau} u_{i})^{phys} \rangle$, which reads

\begin{eqnarray}
i \rho \partial_{i} \langle (\partial_{\tau} u_{i})^{phys} \rangle & = &
\partial_{i} \langle p_{i} \rangle \nonumber \\
& = & - \int \frac{d{\bf k}}{(2 \pi)^{3}} e^{i\bf{k} \cdot \bf{x}}
k^{2} \langle F^{L'} \rangle \; ,
\end{eqnarray}
we see that $\partial_{\tau} \langle (\partial_{i} u_{i})^{phys} \rangle =
\partial_{i} \langle (\partial_{\tau} u_{i})^{phys} \rangle $ if

\begin{eqnarray}
\label{relation a b}
M_{1j} J_{j}  & = & \sqrt{2} a^{L} k J^{L'} \nonumber \\
& = & - \sqrt{2} a^{L}\partial_{\tau}
N_{\Delta}\; .
\end{eqnarray}
Furthermore, we take the interaction with the transverse part of the defect
current density to be zero. This is justified by noting that the
transverse part of the defect density is not a hydrodynamic variable.
We will come back to this point in more detail below.
The total interaction is now uniquely defined in terms of one parameter
$a^{L}$ and given by

\begin{equation}
S_{int}[\vec{F}, \vec{\chi}'] = i \int_{0}^{\hbar \beta}
d\tau \int \frac{d{\bf k}}{(2 \pi)^{3}}
\left\{
a^{L} (k^{2} \chi^{L} + \sqrt{2} \partial_{\tau} F^{L'}) N_{\Delta} \right\}
\; .
\end{equation}

We have now completely specified
the interaction of the point defects with the
phonon field. However, only the phonon field has dynamics up to now.
Because of their interaction with the phonons the point defects
would of course effectively acquire dynamics, but one also expects the
point defects to behave as dynamical particles if one could freeze out
the phonon field. Roughly speaking, they would behave as
particles in a periodic potential that could
tunnel from one minimum to another.
Therefore we have to add a dynamical term for the point defects.
The most general dynamical term that describes propagating behavior
of the defects is

\begin{equation}
\label{free action point defects}
S_{0}[\{ {\bf x}^{(n)} \}] = - \frac{1}{2}
\int_{0}^{\hbar \beta} d\tau   \sum_{n}
x_{i}^{(n)} (m_{ij} \partial_{\tau}^{2}) x_{j}^{(n)} \; .
\end{equation}
The form of the anisotropic mass
$m_{ij}$ is constrained by the symmetry of the crystal.
The field equations are now found by varying with respect to
$\vec{F}$, $\vec{\chi}$ and the positions ${\bf x}^{(n)}$ of the defects,
and are given by

\begin{eqnarray}
\partial_{\tau}^{2} \vec{F} & = & \frac{k^{2}}{2 \rho} P \cdot
\left( A' \cdot \vec{F} - i \frac{2 \rho}{k^{2}}
\sqrt{2} \vec{a} \partial_{\tau}
N_{\Delta} \right) - i Q^{\dagger} \cdot \vec{a} \partial_{\tau} N_{\Delta}
\nonumber \\
\partial_{\tau} \vec{\chi} & = & \frac{1}{2 \rho} Q \cdot
\left( A'  \vec{F} - i \frac{2 \rho}{k^{2}} \sqrt{2} \vec{a} \partial_{\tau}
N_{\Delta} \right) - \frac{i}{k^{2}} R \cdot \vec{a} \partial_{\tau}
N_{\Delta} \\
\partial_{\tau}^{2} m_{ij} x_{j}^{(n)} & = & - i a^{L} q^{(n)} \left(
\sqrt{2} \partial_{\tau}
\partial_{i} F^{L'}
|_{{\bf x} = {\bf x}^{(n)}} -
\partial^{2} \partial_{i} \chi'^{L}|_{{\bf x} =
{\bf x}^{(n)}}
\right) \nonumber \; .
\end{eqnarray}
The equations of motion for ${\bf x}^{(n)}$ can also be written as

\begin{equation}
\partial_{\tau}^{2} x^{(n)}_{i} = - i q^{(n)}
m_{ij}^{-1} \partial_{j} B_{kl} \sigma_{kl}|_{{\bf x} = {\bf x}^{(n)}} \; ,
\end{equation}
where

\begin{equation}
\label{the expression for B}
B_{ij} = \frac{a^{L}}{\sqrt{2}} \left( \delta_{ij} +
\frac{\partial_{i} \partial_{j}}{\partial^{2}} \right) \; .
\end{equation}
This way of writing it will prove useful when deriving the hydrodynamic
equations of motion in Sec. \ref{2C}.
\subsection{Microscopic picture}
\label{2B}
An alternative approach to the derivation of an action which describes
the coupled dynamics of the point defects and the phonons
is to start from a microscopic action.
It describes the atoms constituting the crystal
by their positions

\begin{equation}
{\bf y}^{(i)} = {\bf n}^{(i)} + {\bf u}^{(i)}
\end{equation}
relative to the sites $\{ {\bf n}^{(i)} \}$ of an ideal reference lattice
and assumes an isotropic, short range
interaction $V(|{\bf y}^{(i)} - {\bf y}^{(j)}|)$ between
the individual atoms.
In this approach, it is clear that the hydrodynamic
momentum density is ${\bf g} =  i \rho \partial_{\tau} {\bf u}$
and, as we will see below, what approximations we implicitly made
when we wrote down the free action
of a point defect in Eq. (\ref{free action point defects}).
In first instance, the microscopic action reads

\begin{equation}
S[{\bf u}] = \int_{0}^{\hbar \beta} d\tau \left\{ \frac{1}{2} \sum_{i}
\rho \left( \partial_{\tau} {\bf u}^{(i)} \right)^{2} + \sum_{i < j}
V(|{\bf y}^{(i)} - {\bf y}^{(j)}|) \right\} \; .
\end{equation}
To explicitly include the point defects,
we then decompose the displacement field into a
part describing the phonons and a part describing the defects

\begin{equation}
{\bf u}^{(i)} = {\bf u}^{(i),ph} + {\bf u}^{(i),def} \; ,
\end{equation}
where the defects are located at the positions $\{ {\bf x}^{(n)}(\tau) \}$.
Inserting this decomposition of the displacement field
into the action we get

\begin{eqnarray}
\label{microscopic action}
S[{\bf u}] = \int_{0}^{\hbar \beta} d\tau \left\{ \frac{1}{2} \sum_{i}
\rho \left( \partial_{\tau} {\bf u}^{(i), ph} +
\partial_{\tau} {\bf u}^{(i), def} \right)^{2} +
\sum_{i < j} V(|{\bf y}^{(i)} - {\bf y}^{(j)}|) \right\} \; .
\end{eqnarray}
Since the
positions ${\bf n}^{(i)}$ correspond to the equilibrium positions
of the crystal, the total potential
$V(\{ {\bf x}^{(i)} \})
\equiv \sum_{i < j} V(|{\bf x}^{(i)} - {\bf x}^{(j)}|)$
satisfies

\begin{equation}
\delta V(\{ {\bf n}^{(i)} + {\bf u}^{(i)} \}) =
V(\{ {\bf n}^{(i)} \}) +
{\cal O}[({\bf u}^{(i)})^{2}]
\; .
\end{equation}
For slowly varying displacements, the
quadratic terms equals $(1/2) \int d {\bf x} u_{ij} c_{ijkl} u_{kl}$.
However, we cannot use this long wavelength result to find the interaction
between the phonons and the defects
because defects cause fluctuations in $\bf{u}$ on
the scale of a few lattice spacings. To proceed we therefore
assume that ${\bf u}^{def}$ only changes
significantly over a distance which is much smaller than the typical
wavelength of a phonon.
Moreover, in the continuum limit we can always write for the displacement
of a static defect

\begin{equation}
u_{i}^{def}({\bf y}) = \partial_{i} f({\bf y}) \; ,
\end{equation}
because a defect is defined by a non-zero value of
$v_{0} = \int d{\bf x} u_{ii}$, i.e. the volume that is removed from or
added to the crystal due to the presence of a defect.
The function $f$ thus satisfies $\int d{\bf y} \partial^{2} f = q v_{0}$,
where q is either $-1$ or $+1$ depending on
whether we are dealing with an interstitial or a vacancy.
Note that in the isotropic case
$f({\bf y} - {\bf x}) \propto |{\bf y} -
{\bf x}|^{-1}$, where ${\bf x}$ is again the location of the defect.
Furthermore, to first order in the velocity we can write for a moving
point defect

\begin{equation}
\partial_{\tau} u_{i}^{def}({\bf y}, \tau) \approx
\partial_{\tau} u_{i}^{def}({\bf y} - {\bf x}(\tau)) =
- \partial_{\tau} x_{j} \partial_{j} u_{i}^{def} \; .
\end{equation}
Expanding the action in Eq. (\ref{microscopic action}) up to second order in
the displacements and making use of the above results
the effective action describing phonons and point defects is found to be

\begin{eqnarray}
S[u_{i}] & = & \int_{0}^{\hbar \beta} \int d{\bf x} \left\{
\frac{\rho}{2} (\partial_{\tau} u_{i}^{ph})^{2} +
\frac{1}{2} u_{ij}^{ph} c_{ijkl} u_{kl}^{ph}
\right\} + \nonumber \\
& & \sum_{n} \int d\tau \left\{
- \frac{v_{0}^{2} \rho}{2} x^{(n)}_{i}(\partial_{\tau} m_{ij} \partial_{\tau})
x^{(n)}_{j} +
q^{(n)} v_{0} \tilde{B}_{ij}
u_{ij}^{ph}({\bf x}^{(n)},\tau) \right. \nonumber \\
& & \left. + q^{(n)} v_{0}
\rho [\partial_{\tau} x_{i}^{(n)}] \partial_{i} \partial_{j}
[\partial_{\tau}
u_{j}^{ph}({\bf x}^{(n)},\tau)] + E_{c}
\right\} \; ,
\end{eqnarray}
where we have neglected contributions
with $m \not= n$, assuming the defects to be sufficiently far apart to
interact only through the phonon field. Furthermore, $E_{c}$ denotes the energy
associated with the creation of a defect.
The microscopic action gives certain relations between
the coefficients in this action. However, renormalisation
changes these coefficients and we believe that it
does not preserve the relations between them.
Therefore they have to be treated as independent.
This is important when trying to establish a connection
with the action in terms of
the stress tensor, as found in the previous section,
which can be achieved by means of two
Hubbard-Stratonovich transformations and following the
same route as before by introducing the gauge fields.
The result turns out  to be identical and shows in particular
that there is indeed only an interaction between the phonons and
the longitudinal part of the defect current density.
\subsection{Hydrodynamics }
\label{2C}
We can now derive the hydrodynamic equations
for a crystal with point defects. The number of hydrodynamic
modes is fundamentally related to the number of conserved quantities
and the number of broken symmetries.
The conserved quantities are the total
mass, the total momentum and the netto number of defects $N_{\Delta}$,
the difference between the number of interstitials and vacancies.
The associated conservation laws result in 5 hydrodynamic modes.
In principle we also need to take into account energy
conservation, which would yield an additional
thermal diffusion mode.\cite{Zippelius}
However, for our purposes it is relatively unimportant
and we will not consider it here. Note however that we can
obtain the Hamiltonian from the action and therefore in principle
also include this mode into our considerations.
In addition to the conservation laws,
translational symmetry is spontaneously broken, which results in
3 hydrodynamic Goldstone modes. Hence we expect to find a total of
$8$ hydrodynamic modes.
To find the equations of motion describing these modes we
first identify the hydrodynamic momentum density
$g_{i}$  with $\langle p_{i} \rangle = i \rho
\langle (\partial_{\tau} u_{i})^{phys} \rangle$, which is
obvious from a microscopic point of view because locally it is just the
momentum of the particles of mass $m$ situated on the latice sites.
It is important to note that
it includes the momentum of the point defects, because we
have constructed the physical quantity $(\partial_{\tau} u_{i})^{phys}$
that way. In the remainder of this article we do not explicitly
include the averaging brackets, since it unneccessarily
complicates the notation.

We can immediatly write down the following equality

\begin{equation}
\partial_{\tau} g_{i} = \partial_{j} \sigma_{ij} \; .
\end{equation}
This equation is nothing but the constraints
found in Sec. \ref{2A} in Eq. (\ref{constraints}).
In a perfect crystal without defects, the hydrodynamic
modes are the phonon modes, and their equation of motion
is found by taking the time derivative of the above equation

\begin{equation}
\partial_{\tau}^{2} g_{i} = \partial_{j} \partial_{\tau} \sigma_{ij} \; .
\end{equation}
Therefore we need to know $\partial_{\tau} \sigma_{ij}$ which is easily
calculated as

\begin{eqnarray}
\label{expression for sigma}
\partial_{\tau} \sigma_{ij} & = & \int
\frac{d {\bf k}}{(2 \pi)^{3}}
e^{i {\bf k} \cdot {\bf x}}
\left(
\vec{e}_{ij}^{(1)} \cdot \partial_{\tau}^{2} \vec{F} +
k^{2} \vec{e}_{ij}^{(2)} \cdot \partial_{\tau} \vec{\chi}'
\right) \nonumber \\
& = &
\int \frac{d {\bf k}}{(2 \pi)^{3}} e^{i {\bf k} \cdot {\bf x}}
\left\{ \frac{k^{2}}{2 \rho}
c_{ijkl} \vec{e}_{kl}^{(1)} \cdot A' \cdot \vec{F} -
i c_{ijkl} (\vec{e}_{kl}^{(1)} \cdot
\sqrt{2} \vec{a} + \vec{e}_{kl}^{(2)} \cdot \vec{a})
\partial_{\tau} N_{\Delta} \right\} \nonumber \\
& = &
\int \frac{d {\bf k}}{(2 \pi)^{3}} e^{i {\bf k} \cdot {\bf x}}
\left\{ -\frac{i k_{k}}{\rho} c_{ijkl} g_{l} -
i c_{ijkl} (\sqrt{2} \vec{e}_{kl}^{(1)} \cdot
\vec{a} + \vec{e}_{kl}^{(2)} \cdot \vec{a})
\partial_{\tau} N_{\Delta} \right\} \nonumber \\
& = &
- c_{ijkl}
\left( \frac{1}{\rho} \partial_{k} g_{l} - B_{kl} \partial_{i} J_{i}
\right) \; ,
\end{eqnarray}
where we used the equations of motion for
$\vec{F}$ and $\vec{\chi}'$, the continuity equation
$\partial_{\tau} N_{\Delta} = i \partial_{i} J_{i}$, the
expression for $B_{ij}$ in Eq.
(\ref{the expression for B}) and the following equality

\begin{eqnarray}
\frac{k^{2}}{2} c_{ijkl} \vec{e}_{kl}^{(1)} \cdot A' \cdot \vec{F}
& = & k^{2} c_{ijkl} \hat{k}
\left(
\frac{e_{l}^{(1,1)}}{\sqrt{2}} F^{(2,1)} +
\frac{e_{l}^{(1,-1)}}{\sqrt{2}} F^{(2,-1)} + e_{l}^{(1,0)} F^{L'}
\right) \nonumber \\
& = & k^{2} c_{ijkl} \hat{k}_{k} \hat{k}_{m} F_{ml} \nonumber \\
& = & - i k_{k} c_{ijkl} i k_{m} F_{ml} \nonumber \\
& = & - i k_{k} c_{ijkl} g_{l} \; .
\end{eqnarray}
Thus we find for $g_{i}$ the folowing equation of motion,
describing the phonon modes
and their interaction with the point defects

\begin{equation}
\label{hydr.eq. point defects}
\partial_{\tau}^{2} g_{i}  =  - c_{ijkl} \partial_{j} \left(
\partial_{k} \frac{g_{l}}{\rho} - B_{kl} \partial_{m} J_{m}
\right) \; .
\end{equation}
We can split the above equation into three continuity equations as follows

\begin{eqnarray}
\label{1}
\partial_{\tau} \delta \rho & = & i \partial_{i} g_{i} \nonumber \\
\partial_{\tau} \vartheta_{i} & = & i \varepsilon_{ijk} \partial_{j} g_{k} \\
\partial_{\tau} g_{i} & = & i c_{ijkl} \partial_{j}
\left\{
\frac{\partial_{k}}{\rho}
\left(
\frac{\partial_{l} \delta \rho}{\partial^{2}} +
\frac{\varepsilon_{lmn} \partial_{m} \vartheta_{n}}{\partial^{2}}
\right)
- B_{kl} N_{\Delta}
\right\} \; . \nonumber
\end{eqnarray}
Note that it follows from Eq. (\ref{1}), combined with the continuity
equation for $g_{i}$, that $\sigma_{ij}$ can be written as

\begin{equation}
\label{expression sigma}
\sigma_{ij} = - i c_{ijkl}
\frac{\partial_{k}}{\rho}
\left(
\frac{\partial_{l} \delta \rho}{\partial^{2}} +
\frac{\varepsilon_{lmn} \partial_{m} \vartheta_{n}}{\partial^{2}}
\right)
- i B_{kl} N_{\Delta} \; .
\end{equation}

To find the total of $8$ modes, instead of the $6$ given by the equations
above, we need to include point defects into our hydrodynamic equations, as
was first noted by Martin {\it et.al.}.\cite{Martin} Thus, we have an
additional hydrodynamic variable, the netto defect density $N_{\Delta}$.
Note that the transverse part of the
defect current density is not a hydrodynamic variable
because the momentum of the defects is not conserved.
We now want to write down the equation of motion for $N_{\Delta} = \sum_{n}
q ^{(n)} \delta({\bf x} - {\bf x}^{(n)})$.
Using the equations of motion for the point defects, we get

\begin{eqnarray}
\label{defect density hydrodynamics}
\partial_{\tau}^{2} \sum_{n} \delta({\bf x} - {\bf x}^{(n)})
& = &
\sum_{n} q^{(n)} \left\{
\dot{x}^{(n)}_{i} \dot{x}^{(n)}_{j} \partial_{i} \partial_{j}
\delta({\bf x} - {\bf x}^{(n)}) - \ddot{x}^{(n)}_{i} \partial_{i}
\delta({\bf x} - {\bf x}^{(n)}) \right\} \nonumber \\
& = &
\sum_{n} q^{(n)} \left\{
\dot{x}^{(n)}_{i} \dot{x}^{(n)}_{j} \partial_{i} \partial_{j}
\delta({\bf x} - {\bf x}^{(n)}) + i m_{ij}^{-1} \partial_{j} B_{kl} \sigma_{kl}
\partial_{i} \delta({\bf x} - {\bf x}^{(n)}) \right\} \; .
\end{eqnarray}
To find the hydrodynamics of the netto defect density $N_{\Delta}$ we have to
average Eq. (\ref{defect density hydrodynamics}) over the initial
conditions. In an isotropic gas in the absence of external forces, the term
proportional to $\sum_{n} \dot{x}_{i}^{(n)} \dot{x}_{j}^{(n)}
\delta({\bf x} - {\bf x}^{(n)})$ would be the only term present,
the average of which is just the pressure tensor $\pi_{ij}$.
A closed set of equations
giving the linearized hydrodynamics would then be found by
writing down a gradient
expansion for $\pi_{ij}$ in terms of the hydrodynamic variables.
In our case to lowest order the pressure tensor
can only be a function of $\delta N_{\Delta} \equiv N_{\Delta} -
\angle N_{\Delta} \rangle$  because
it is the only variable which is even under time reversal, and we get
$\pi_{ij} = M_{ij} \delta N_{\Delta} + {\cal O}(\delta N_{\Delta}^{2})$.
Neglecting the terms quadratic in the fluctuations
we find for $\delta N_{\Delta}$ the following linearized
equation of motion

\begin{equation}
\label{hydrodynamic equation 2}
\partial_{\tau}^{2} N_{\Delta} = - M_{ij} \partial_{i} \partial_{j}
N_{\Delta} + i m_{ij}^{-1} \partial_{i} \partial_{j} B_{kl} \sigma_{kl} \; ,
\end{equation}
where $B_{kl} \rightarrow B_{kl} \langle N_{\Delta} \rangle$.
Note that if we had naively introduced the dynamics of the defects
into our theory by adding a Lagrangian density for the defect density
instead of the defects
${\cal L} = - \frac{1}{2} N_{\Delta}([M_{ij} \partial_{i} \partial_{j}]^{-1}
\partial_{\tau}^{2} + 1)N_{\Delta}$
as was done by Stoof {\it et.al.},\cite{Stoof} we would
have obtained the same equation by varying
the action with respect to $N_{\Delta}$.
However, our approach is more fundamental and illumates clearly
the underlying physics of this Lagrangian density.
The hydrodynamic equations are usually given as a set of continuity
equations, i.e. with first order time derivatives.
Therefore we rewrite Eq. (\ref{hydrodynamic equation 2})
as a pair of continuity equations

\begin{eqnarray}
\label{4.2}
\partial_{\tau} N_{\Delta} & = & i \partial_{i} J_{i} \nonumber \\
\partial_{\tau} J_{i} & = & i \partial_{i}\left(
\frac{M_{ij} \partial_{i} \partial_{j}}{\partial^{2}}  N_{\Delta} -
i \frac{m^{-1}_{ij} \partial_{i} \partial_{j}}{\partial^{2}}
B_{kl} \sigma_{kl} \right) \; .
\end{eqnarray}
We stress that this is actually only an equation for the
longitudinal part of the defect current,
$J_{i}^{L} = \partial_{i} \partial_{j}/\partial^{2} J_{j}$.
The transverse part is not a hydrodynamic variable and
is anticipated to relax to zero on a microscopic time scale.
This completes our discussion of the dissipationless hydrodynamic equations.
We have obtained a set of hydrodynamic equations describing
phonons, point defects and their interaction for a HCP crystal.
They are given by Eqs. (\ref{1}) and (\ref{4.2}).

It is interesting to note that these equations can also be derived from a
hydrodynamic action of which the Lagrangian density is given by

\begin{eqnarray}
\label{hydrodynamic action}
{\cal L} = \frac{1}{2} \sigma_{ij}
\left\{
\frac{\partial_{j} \partial_{l} \delta_{ik}}{\rho \partial_{\tau}^{2}} +
c_{ijkl}^{-1}
\right\} \sigma_{kl} + i \sigma_{kl} B_{kl} N_{\Delta}
- \frac{1}{2} N_{\Delta}
\left\{
\frac{\partial_{\tau}^{2}}{m_{ij} \partial_{i} \partial_{j}} + E_{c}
\right\}
N_{\Delta} \; ,
\end{eqnarray}
using $M_{ij} \partial_{i} \partial_{j} =
E_{c} m_{ij}^{-1} \partial_{i} \partial_{j}$.
{}From the associated action, the hydrodynamic equations
describing the phonon modes and their coupling to $N_{\Delta}$
are found by writing down the field equations for
$\sigma_{ij}$ and defining $g_{i}$ by the
constraint $\partial_{\tau} g_{i} = \partial_{j} \sigma_{ij}$.
Note that in Eq. (\ref{hydrodynamic action})
the term quadratic in $\sigma_{ij}$ is just
$p_{i}^{2}/2 \rho + \sigma_{ij} c_{ijkl}^{-1} \sigma_{kl}$,
which is the free part of the action in Eq. (\ref{action 100})
with the substition
$p_{i} \rightarrow \partial_{j} \sigma_{ij}/\partial_{\tau}$.
Therefore, Eq. (\ref{hydrodynamic action}) is the analogue of
the hydrodynamic action describing density
fluctuations in a normal fluid.\cite{Forster,Popov}

For completeness we write down
the total set of hydrodynamic equations which as expected
amount to a total of 8 continuity equations

\begin{eqnarray}
\partial_{\tau} \delta \rho & = & i \partial_{i} g_{i} \nonumber \\
\partial_{\tau} \vartheta_{i} & = & i \varepsilon_{ijk} \partial_{j} g_{k}
\nonumber \\
\partial_{\tau} g_{i} & = & i c_{ijkl} \partial_{j}
\left\{
\frac{\partial_{k}}{\rho}
\left(
\frac{\partial_{l} \delta \rho}{\partial^{2}} +
\frac{\varepsilon_{lmn} \partial_{m} \vartheta_{n}}{\partial^{2}}
\right)
- B_{kl} N_{\Delta}
\right\} \\
\partial_{\tau} N_{\Delta} & = & i \partial_{i} J_{i} \nonumber \\
\partial_{\tau} J_{i} & = & i \partial_{i}\left(
\frac{M_{ij} \partial_{i} \partial_{j}}{\partial^{2}}  N_{\Delta} -
\frac{i m^{-1}_{ij} \partial_{i} \partial_{j}}{\partial^{2}}
B_{kl} \sigma_{kl} \right) \nonumber \; .
\end{eqnarray}
To check heuristically if we ended up with the right equations
we write down the hydrodynamic equations
in the case of a isotropic crystal and compare these
to the ones found for a
two-dimensional isotropic crystal by Stoof {\it et.al.}\cite{Stoof}
In the isotropic case $c_{ijkl}$, $B_{ij}$, $M_{ij}$ and $m_{ij}$
are given by

\begin{eqnarray}
c_{ijkl} & = &\lambda \delta_{ij} \delta_{kl} + \mu
( \delta_{ik} \delta_{jl} + \delta_{il} \delta_{jk} ) \nonumber \\
M_{ij} & = & M \delta_{ij} \\
m_{ij} & = & m \delta_{ij} \nonumber \; .
\end{eqnarray}
This implies the following equalities

\begin{eqnarray}
c_{ijkl} \partial_{j} \partial_{k} \partial_{l}
\frac{\delta \rho}{\rho \partial^{2}} & = &
\frac{(\lambda + 2 \mu)}{\rho} \partial_{i} \delta \rho \nonumber \\
c_{ijkl} \partial_{j} \partial_{k} \varepsilon_{lmn} \partial_{m}
\frac{\theta_{n}}{\rho \partial^{2}} & = &
\frac{\mu}{\rho} \varepsilon_{imn} \partial_{m} \theta_{n} \\
c_{ijkl} B_{kl} \partial_{j}
N_{\Delta} & = & \frac{4}{\sqrt{2}}
(\mu + \lambda) a^{L} \partial_{i} N_{\Delta} \nonumber \; .
\end{eqnarray}
Furthermore, we write the hydrodynamic equations in real time, which
amount to the substitution $\partial_{\tau} \rightarrow -i \partial_{t}$.
As a result the hydrodynamic equations for an isotropic three-dimensional
crystal with point defects are given by

\begin{eqnarray}
\label{stoof1}
\partial_{t} \delta \rho & = & - \partial_{i} g_{i} \nonumber \\
\partial_{t} \vartheta_{i} & = & - \varepsilon_{ijk} \partial_{j} g_{k}
\nonumber \\
\partial_{t} g_{i} & = &
- \left\{
\frac{\lambda + 2\mu}{\rho} \partial_{i} \delta \rho +
\frac{\mu}{\rho} \varepsilon_{ikl} \partial_{k} \theta_{l} -
\frac{4}{\sqrt{2}}(\mu + \lambda) a^{L} \partial_{i} N_{\Delta}
\right\} \\
\partial_{t} N_{\Delta} & = & - \partial_{i} J_{i} \nonumber \\
\partial_{t} J_{i} & = & - c_{\rho} \partial_{i} \delta \rho -
c_{\Delta} \partial_{i} N_{\Delta} \nonumber \; ,
\end{eqnarray}
where $c_{\rho} = [m a^{L}/\sqrt{2}](4 \lambda + 4 \mu)$ and
$c_{\Delta} = M - 2 m (a^{L})^{2}(4 \lambda + 3 \mu)$.
These are indeed the three-dimensional generalisations
of the equations found by Stoof {\it et.al.} for
the two-dimensional isotropic cristal without dissipation.

In order to give a realistic description of the system, we need to
include dissipational effects into our hydrodynamic equations.
Although there is a coupling between the phonon
field and the defect density and thus a `shake up' of the phonon field if a
defect moves, up to this point there is no real dissipation because
the bilineair coupling between the phonon
and the defect modes causes mixing
of these modes, but no dissipation. Therefore we
include dissipation into our
hydrodynamic equations in the standard way by first
expanding the dissipative parti of the stress tensor
 to linear order in the conjugate forces and
requiring the coefficients
to be compatible with the symmetry of the system under consideration
and then in turn expanding the conjugate forces in terms of the
currents.\cite{Liu,Martin,Putterman} A particularly clear treatment
of this standard method is given by F. J\"ahnig and H. Schmidt.\cite{Jahnig}
Quite generally, our hydrodynamic equations have the following form

\begin{equation}
\begin{array}{lcl}
\partial_{t} \delta \rho & = & - \partial_{i} g_{i} \\
\partial_{t} \theta_{i} & = & - \varepsilon_{ijk} \partial_{j} g_{k} \\
\partial_{t} g_{i} & = & - \partial_{j}(\sigma_{ij} + \sigma_{ij}^{D}) \\
\partial_{t} N_{\Delta} & = & - \partial_{i} J_{i} \\
\partial_{t} J_{i} & = & - \partial_{j} (\pi_{ij} +
\pi_{ij}^{D}) \; ,
\end{array}
\end{equation}
where the superscript $D$ denotes the dissipative part of the
`stress' tensors, and the non-dissipative part has already been determined.
Roughly speaking
the variables $\theta_{i}$ and $J_{i}$ are associated with the
3 broken symmetries, whereas $\rho$, $g_{i}$, $N_{\Delta}$ account for
the conservation of mass, momentum and defects.
The most general dissipative terms allowed by the requirement
that the time reversal symmetry of the dissipative currents
is opposite to the associated hydrodynamic variable are given by

\begin{eqnarray}
\sigma_{ij}^{D} & = & \frac{1}{\rho} \eta_{ijkl}^{(1)} \partial_{k} g_{l}
+ \zeta^{(1)}_{ij} \partial_{k} J_{k} \nonumber \\
\pi_{\Delta, ij}^{D} & = &
\frac{1}{\rho} \delta_{ij} \eta_{kl}^{(2)}
\partial_{k} g_{l}
+ \delta_{ij} \zeta^{(4)} \partial_{k} J_{k} \; ,
\end{eqnarray}
where we used that $J_{i}$ containes only a longitudinal degree of
freedom. The specific form of the parameters is determined by the
discrete symmetries of the system, which in the case of the
HCP crystal $^{4}He$ form the group ${\cal C}_{6h}$.
It should be noted that in Stoof {\it et.al.} it was incorrectly assumed
that the transverse part of the
defect current behaves as in a gas and diffuses to zero.\cite{Stoof}
As we have seen, the correct behavior of the transverse part
of the defect current is a relaxation to zero on a microscopic time scale.
\section{SUPERSOLID HYDRODYNAMICS}
\label{section3}
In view of the exciting experiments by Lengua and Goodkind,
our aim in writing this paper was also to formulate the hydrodynamic
equations of supersolid $^{4}$He.
Hence we have come to the point where we have to include into our
hydrodynamic equations the superfluid degree of freedom.
{}From microscopic theories developed for superfluid
liquids and gases it is well known
how we should proceed to include these additional
degrees of freedom into the hydrodynamic equations for the normal
phase.\cite{Hohenberg,Kirkpatrick}
First, the density $\rho$ is split into a normal part $\rho^{n}_{ij}$ and
a superfluid part $\rho^{s}_{ij}$, satisfying

\begin{equation}
\rho \delta_{ij} =
\rho^{s}_{ij} + \rho^{n}_{ij} \; .
\end{equation}
Note that the tensorial nature of the
densities is of importance in the case of an
anisotropic HCP crystalline structure.
Second, we split the total momentum density of the system
into a normal part $\rho^{n}_{ij}v^{n}_{j}$ and a superfluid part
$\rho^{s}_{ij}v^{s}_{j}$ according to

\begin{eqnarray}
g_{i} & = & \rho_{ij}^{s} v^{s}_{j} + \rho_{ij}^{n} v_{j}^{n} \nonumber \\
& = & \rho v_{i}^{n} + \rho^{s}_{ij}(v^{s}_{j} - v^{n}_{j}) \; ,
\end{eqnarray}
where the superfluid velocity is purely
longitudinal, i.e. $\varepsilon_{ijk} \partial_{j} v^{s}_{k} = 0$,
because it is proportional to the gradient of the superfluid
phase $\phi^{s}$.

Furthermore, the dissipative terms have to be generalized for an
anisotropic superfluid, and the
dynamics of the superfluid velocity has to be determined.
Following the standard treatment,
the dissipative part of the stress tensor $\sigma_{ij}^{D}$ becomes

\begin{equation}
\sigma_{ij}^{D} =
\eta^{(1)}_{ijkl} \partial_{k} v_{l}^{n} +
\zeta^{(1)}_{ij} \partial_{k} J_{k} +
\frac{1}{\rho} \zeta_{ij}^{(3)} \partial_{k}
\rho_{kl}^{s} (v_{l}^{s} - v_{l}^{n}) \; .
\end{equation}
The last term on the right-hand side
is the most general term containing the conjugate variable of
the phase field, i.e.
$\partial_{i} g_{i}^{s} \equiv \partial_{i} \rho_{ij}
(v_{j}^{n} - v_{j}^{s})$.\cite{Liu,Martin,Putterman}
Furthermore, the dynamics of the superfluid phase field
is basically determined by the Josephson
relation and is given by

\begin{eqnarray}
\label{superfluid dynamics}
\partial_{t} v^{s}_{i}  & = &
- \frac{B_{\rho}}{\rho^{2}} \partial_{i} \delta \rho +
\beta_{\Delta} \partial_{i} N_{\Delta} +
\partial_{i} \zeta^{(7)}_{jk} \partial_{j} v_{k}^{n} +
\zeta^{(8)} \partial_{i} \partial_{j} J_{j} +
\frac{\zeta^{(10)}}{\rho} \partial_{i} \partial_{j}
\rho_{jk}^{s} (v_{k}^{s} - v_{k}^{n}) \; ,
\end{eqnarray}
where $B_{\rho} = \rho^{2} (\partial \mu /
\partial \rho)|_{T, n_{\Delta}}$
is the isothermal bulk
modulus, $\mu$ is the chemical potential per unit mass
and $\beta_{\Delta} =
- \partial \mu / \partial n_{\Delta}|_{\rho, T}$.
By adding the last three terms in the right-hand side
of  Eq. (\ref{superfluid dynamics}) we
have also included dissipation.
However, at this point we
have to realize that we were already dealing with a two
fluid hydrodynamics in the normal solid phase,
due to the presence of defects.
This means that we also have to split
the defect current density $J_{i}$ into a normal and a superfluid part, i.e.
$J_{i} = J_{i}^{n} + J_{i}^{s}$.
Physically, this means that the
superfluid current density can be caused
both by the motion of defects, and by lattice vibrations.\cite{Stoof}
As a result we end up with the following hydrodynamic equations describing
supersolid $^{4}$He

\begin{eqnarray}
\partial_{t} \delta \rho & = & - \partial_{i} g_{i} \nonumber \\
\partial_{t} \theta_{i} & = & - \varepsilon_{ijk} \partial_{j} g_{k}
\nonumber \\
\partial_{t} g_{i} & = & - c_{ijkl} \partial_{j}
\left\{
\frac{\partial_{k}}{\rho}
\left(
\frac{\partial_{l} \delta \rho}{\partial^{2}} +
\frac{\varepsilon_{lmn} \partial_{m} \vartheta_{n}}{\partial^{2}}
\right) - B_{kl} N_{\Delta}
\right\} + \nonumber \\
& & \eta^{(1)}_{ijkl} \partial_{j} \partial_{k} v_{l}^{n} +
\zeta^{(1)}_{ij} \partial_{j} \partial_{k} J_{k}^{n}  +
\zeta^{(2)}_{ij} \partial_{j} \partial_{k} (J_{k}^{s} - J_{k}^{n}) +
\frac{1}{\rho} \zeta_{ij}^{(3)} \partial_{k}
\rho_{kl}^{s} (v_{l}^{s} - v_{l}^{n}) \nonumber \\
\partial_{t} N_{\Delta} & = & - \partial_{i} J_{i} \\
\partial_{t} J_{i} & = &  - \partial_{i} \left(
\frac{\partial_{i} \partial_{j} M_{ij}}{\partial^{2}}
N_{\Delta} - \frac{\partial_{i} \partial_{j} m_{ij}^{-1}}{\partial^{2}}
\partial_{j} B_{kl} i \sigma_{kl} \right) + \nonumber \\
& & \eta_{jk}^{(2)} \partial_{i} \partial_{j} v_{k}^{n}
+ \zeta^{(4)} \partial_{i} \partial_{j} J_{j}^{n} +
\zeta^{(5)} \partial_{i} \partial_{j} (J_{j}^{s} - J_{j}^{n}) +
\frac{1}{\rho} \zeta^{(6)} \partial_{i} \partial_{j}
\rho_{jk}^{s} (v_{k}^{s} - v_{k}^{n})  \nonumber \\
\partial_{t} v^{s}_{i}  & = &
- \frac{B_{\rho}}{\rho^{2}} \partial_{i} \delta \rho +
\beta_{\Delta} \partial_{i} N_{\Delta} + \nonumber \\
& & \partial_{i} \zeta^{(7)}_{jk} \partial_{j} v_{k}^{n} +
\zeta^{(8)} \partial_{i} \partial_{j} J^{n}_{j} +
\zeta^{(9)} \partial_{i} \partial_{j} (J_{j}^{s} - J_{j}^{n}) +
\frac{\zeta^{(10)}}{\rho} \partial_{i} \partial_{j}
\rho_{jk}^{s} (v_{k}^{s} - v_{k}^{n}) \nonumber \\
\partial_{t} J^{s}_{i} & = &
- \frac{B_{\Delta}}{\rho^{2}} \partial_{i} N_{\Delta} +
\beta_{\rho} \partial_{i} \delta \rho + \nonumber \\
& & \partial_{i} \zeta^{(11)}_{jk} \partial_{j} v_{k}^{n} +
\zeta^{(12)} \partial_{i} \partial_{j} J^{n}_{j} +
\zeta^{(13)} \partial_{i} \partial_{j} (J_{j}^{s} - J^{n}_{j}) +
\frac{1}{\rho} \zeta^{(14)} \partial_{i} \partial_{j}
\rho_{jk}^{s} (v_{k}^{s} - v_{k}^{n}) \nonumber \; .
\end{eqnarray}
The large number of dissipative terms makes these
equations look rather intricate, but in the limit $k \rightarrow 0$ only the
non-dissipative terms remain and a considerable simplification occurs,
as we will see below.
They are easily seen to
represent ten equations for the ten unknown quantities
$\delta \rho, \theta_{i}, v^{n}_{i}, v^{s}_{i}, N_{\Delta}, J_{i}$ and
$J_{i}^{S}$, realizing that $v^{s}_{i}, J_{i}^{s}$ and
$J_{i}^{n}$ have only one degree of freedom
and $\theta_{i}$ has only two degrees of freedom.
\section{COMPARISON WITH EXPERIMENT}
\label{section4}
We now want to compare our results
with the equations used by Lengua and Goodkind
to fit the data of their experiment
in which they may have observed the supersolid
phase of $^{4}$He.\cite{Lengua}
Their phenomenological equations describe a set of two coupled harmonic
oscillators. We show below that these equations essentially follow
from our hydrodynamic equations describing a normal
crystal with defects.

To find the mode structure present in our
dissipationless hydrodynamic equations,
it is convenient to rewrite Eq.
(\ref{hydrodynamic equation 2}) in terms of the longitudinal
part of the defect current density $J_{i}$.
After taking the time derivative of the second equation
of Eq. (\ref{4.2}) and inserting the first equation we get

\begin{equation}
\partial_{t}^{2} J_{i} = \partial_{i} \left(
\frac{ m_{ij}^{-1} \partial_{i} \partial_{j}}{\partial^{2}}
B_{kl} i \partial_{t} \sigma_{kl} + \frac{M_{ij} \partial_{i} \partial_{j}}
{\partial^{2}} \partial_{k}
J_{k} \right) \; .
\end{equation}
To obtain a closed set of equations
we then use Eq. (\ref{expression for sigma}), which expresses
$\sigma_{ij}$ in terms of $g_{i}$ and $N_{\Delta}$.
We find

\begin{equation}
\label{hydro defect current}
\partial_{t}^{2} J_{i} = \partial_{i} \left[
\frac{m_{ij}^{-1} \partial_{i} \partial_{j}}{\partial^{2}} B_{mn} c_{mnkl}
\left( \frac{1}{\rho} \partial_{k} g_{l} - B_{kl} \partial_{i} J_{i}
\right) + \frac{M_{ij} \partial_{i} \partial_{j}}{\partial^{2}}
\partial_{k} J_{k} \right] \; .
\end{equation}

We now turn to Eq. (\ref{hydr.eq. point defects})
which describes the phonon modes.
First we define the eigenvectors $A^{(n)}_{i}(\hat{\bf k})$,
$n = \{ 1,2,3 \}$, of the matrix $c_{ijkl} k_{j} k_{k}$
as follows

\begin{equation}
\label{eigenvectors2}
c_{ijkl} k_{j} k_{k} A^{(n)}_{l}
(\hat{\bf k})
 = k^{2} \lambda_{n}^{2}(\hat{\bf k}) A_{i}^{(n)}(\hat{\bf k}) \; .
\end{equation}
The 6 phonon modes of the ideal crystal are thus given by
$A_{i}^{(n)}(\hat{\bf k})
e^{i(\omega({\bf k}) t \pm i {\bf k} \cdot {\bf x})}$, with
$\omega^{2}({\bf k}) = k^{2} \lambda_{n}^{2}(\hat{\bf k})$.
In order to find the equations used by Lengua and Goodkind
we first expand $g_{i}$ in terms of the eigenvectors
$A_{i}^{(n)}(\hat{\bf k})$, i.e.
$g_{i} = \sum_{n}g^{(n)}({\bf k}) A_{i}^{(n)}(\hat{\bf k})$.
We then write $J_{i} = J^{L}({\bf k}, \tau) \hat{k}_{i}$
and insert these expressions into Eq. (\ref{hydr.eq. point defects})
and Eq. (\ref{hydro defect current}).
After contracting the first equation with the eigenvectors $A_{i}$ and the
second with $\hat{k}_{i}$,
this leads to the following equations in Fourier space

\begin{eqnarray}
\label{hydrodynamic2}
\partial_{t}^{2} g^{(n)}
& = & - \frac{1}{\rho} k^{2} \lambda^{2}_{n}(\hat{\bf k}) g^{(n)} +
k^{2} \alpha^{(n)}(\hat{\bf k}) J^{L} \nonumber \\
\partial_{t}^{2} J^{L} & = & - m_{ij} k_{i} k_{j}
\left( \beta(\hat{\bf k}) J^{L} +
\frac{1}{\rho}
\sum_{n} \alpha^{(n)}(\hat{\bf k}) g^{(n)} \right) \; ,
\end{eqnarray}
where we defined $\alpha^{(n)}(\hat{\bf k}) \equiv B_{ij} c_{ijkl} \hat{k}_{j}
A_{l}^{(n)}$ and $\beta(\hat{\bf k}) \equiv (M_{ij} k_{i}
k_{j})/(m_{ij} k_{i} k_{j})  - B_{ij} c_{ijkl} B_{kl}$.
Finally we consider one particular mode, say $m$, and elliminate the
two modes with $n \not= m$.
After Fourier transforming also the time variable
the equations for $g^{(n)}$ with $n \not= m$ are solved by

\begin{equation}
g^{(n)} = \frac{k^{2} \alpha^{(n)}(\hat{\bf k}) J^{L}}{\omega^{2} -
\frac{k^{2} \lambda_{n}^{2}}{\rho}} \; .
\end{equation}
Inserting this into Eq. (\ref{hydrodynamic2}) we find

\begin{eqnarray}
- \omega^{2} g^{(m)}
& = & - \frac{1}{\rho} k^{2} \lambda^{2}_{m} g^{(m)} +
k^{2} \alpha^{(m)}(\hat{\bf k}) J^{L} \nonumber \\
- \omega^{2} J^{L} & = & - m_{ij} k_{i} k_{j} \left\{ \beta(\hat{\bf k})
J^{L} + \frac{1}{\rho} \alpha^{(m)}(\hat{\bf k}) g^{(m)} +
\frac{1}{\rho} \sum_{n \not= m}
\frac{\alpha^{(n)}(\hat{\bf k}) \alpha^{(n)}(\hat{\bf k}) k^{2} J^{L}}
{\omega^{2} - \frac{k^{2} \lambda_{n}^{2}}{\rho}}
\right\} \; .
\end{eqnarray}
These equations still contain $4$ separate modes.
However, solutions to these equations have $\omega^{2} \propto k^{2}$.
Therefore we essentially find the following equations

\begin{eqnarray}
\label{two modes 1}
\partial_{t}^{2} g^{(m)}
& = & - \frac{1}{\rho} k^{2} \lambda^{2}_{n} g^{(m)} + k^{2} \alpha^{(m)}
(\hat{{\bf k}}) J^{L} \nonumber \\
\partial_{t}^{2} J^{L} & = & k_{i} m_{ij} k_{j}
\left( \beta'(\hat{\bf k}) J^{L}
+ \frac{\alpha^{(m)}(\hat{{\bf k}})}{\rho} g^{(m)}
\right) \; .
\end{eqnarray}
These indeed describe a set of coupled harmonic oscillators and
agree with the dissipationless limit of the
equations used by Lengua and Goodkind to interprete their data.

If we now add dissipation, the modes $A^{(n)}$
no longer diagonalize Eq. (\ref{hydr.eq. point defects}).
However, there will be
a new set of damped phonon modes with imaginairy eigenvalues.
Proceeding as before, we can again elliminate two modes.
We then find a coupled set of damped harmonic oscillators
that now precisely agree with the equations used by Lengua and Goodkind.

To conclude this section, let us consider the dissipationless hydrodynamic
equations describing an isotropic supersolid. The transverse phonon modes
then decouple, and for the longitudinal
part we find schematically the following equations

\begin{eqnarray}
\label{first and second sound}
  \partial_{t}^{2}
  \left(
    \begin{array}{c}
      \delta \rho \\
      N_{\Delta}
    \end{array}
  \right)
  & = &
  \partial^{2}
  \left(
    \begin{array}{cc}
      \frac{\lambda + 2\mu}{\rho} & -\frac{4}{\sqrt{2}}(\mu + \lambda)a^{L} \\
      c_{\rho} & c_{\Delta}
    \end{array}
  \right)
  \left(
    \begin{array}{c}
      \delta \rho \\
      N_{\Delta}
    \end{array}
  \right)
\nonumber \\
\partial_{\tau}
\left(
  \begin{array}{c}
    v_{i}^{s} \\
    J_{i}^{s}
  \end{array}
\right)
& = &
\partial_{i}
\left(
  \begin{array}{cc}
    - \frac{B_{\rho}}{\rho^{2}} & \beta_{\Delta} \\
    - \frac{B_{\Delta}}{\rho^{2}} & \beta_{\rho}
  \end {array}
\right)
\left(
  \begin{array}{c}
    \delta \rho \\
    N_{\Delta}
  \end{array}
\right) \; .
\end{eqnarray}
The hydrodynamic modes can in principle be found by
diagonalizing the two matrices.
If we are in the normal phase, the first equation remains unchanged,
whereas the second is absent. Clearly we then have four propagating
sound modes. In the supersolid phase the second equation is also present,
and we find two second sound modes in addition to the four first sound modes.
These are however not accurately described
by Eq. (\ref{first and second sound}), because for that
it is essential to include temperature fluctuations,
which we have neglected throughout this article.
Nevertheless, it is clear from the above that to show experimentally the
existence of a supersolid, it would be very convincing
if one observes an additional resonance
due to one of the modes associated with the superfluid degrees of freedom.
\section{CONCLUSION}
\label{section5}
We have derived the hydrodynamic equations
for the solid and supersolid phases of $^{4}$He.
It is well known that to describe the normal solid phase,
it is essential to include defects into the hydrodynamic equations
to find the right number of modes predicted by the conservation laws
and broken symmetries.
Because we know that there are $6$ phonon modes,
the defects are usually assumed to have diffusive dynamics,
giving a total of $6 + 1 = 7$ hydrodynamic modes.
This is then in agreement with the $8 - 1$ modes
one expects from the usual counting argument, excluding
a thermal diffusion mode.
However, Lengua and Goodkind in their experiment observe instead propagating
behavior of the defect mode. This brings the total number of hydrodynamic
modes to $6 + 2 = 8$. Therefore we introduced another
hydrodynamic variable, the longitudinal part of the defect momentum.
We believe that this is justified by noting that, when counting the number
of conserved quantities, we should also include the conservation of defects.
Hence the continuity equations for $N_{\Delta}$ and $J_{i}$
are roughly speaking
associated with respectively a conservation law and a broken symmetry.
Indeed, our equations  reproduce the set of coupled wave equations
which were used by Lengua and Goodkind to interpret their data,
and lead them to the identification of
the observed collective mode as a propagating defect mode.

Furthermore, we have considered the hydrodynamic equations of supersolid
$^{4}$He by allowing both fluctuations in the defects density
and lattice vibrations to lead to superfluid motion.\cite{Stoof}
If we include these superfluid degrees of freedom into
our hydrodynamic equations in the standard way,
we end up with what one might call a four fluid hydrodynamics
instead of the usual two fluid hydrodynamics. As a result we end up with
two second sound modes instead of one. We expect on general grounds
that including temperature fluctuations leads to one of these modes
becoming propagating whereas the other will remain diffusive.
Given these results it should then be possible in principle
to identify experimentally an additional resonance in the attenuation and
velocity of sound due to the coupling of these modes to the phonons.
In our opinion  this would be a more convincing
experimental proof for the existence
of a supersolid phase than the analysis made by Lengua and Goodkind.

\section*{acknowledgements}
This research was supported by the Stichting voor Fundamenteel
Onderzoek der Materie (FOM) which is financially supported by the Nederlandse
Organisatie vorr Wetenschappelijk Onderzoek (NWO). We thank Steve Girvin for
many stimulating and helpful discussions.

\begin{figure}
\caption{Tentative scetch of the phase diagram of $^4$He. 
         \label{fig1}}
\end{figure}


\begin{references}
\bibitem{Keller} W.E.Keller, {\it Helium-3 and Helium-4}
(Plenum Press, New York, 1969)
\bibitem{Landau2} L.D. Landau, J. Phys. U.S.S.R. {\bf 5}, 71 (1941)
\bibitem{Feynman} R.P. Feynman, Phys. Rev. {\bf 91}, 1301 (1953);
Phys. Rev. {\bf 94}, 262 (1954)
\bibitem{Hohenberg} P.C. Hohenberg and P.C. Martin, Ann. Phys. {\bf 34},
291 (1965).
\bibitem{Rudnick} I. Rudnick, Phys. Rev. Lett. {\bf 40}, 1454 (1978);
D.J. Bishop and J.D. Reppy, {\it ibid.} {\bf 40}, 1727 (1978)
\bibitem{Varma} C.M. Varma and N.R. Werthamer, {\it The Physics of Liquid
and Solid Helium} eds. K.H. Bennemann and J.B. Ketterson
(Wiley, New York, 1976).
\bibitem{Meisel} M.W. Meisel, Physica B {\bf 178} 121 (1992) and references
therein.
\bibitem{Andreev} A.F. Andreev and I.M. Lifshits, Sov. Phys. JETP {\bf 29},
1107 (1969).
\bibitem{Chester} G.V. Chester, Phys. Rev. A {\bf 2} 256 (1970)
\bibitem{Leggett2} A.J.Leggett, Phys. Rev. Lett. {\bf 25} 1543 (1970)
\bibitem{Lengua} G.A. Lengua and J.M. Goodkind, J. of Low Temp. Phys. {\bf 79}
, 251 (1990).
\bibitem{Stoof} H.T.C Stoof, K. Mullen, M. Wallin and S.M. Girvin,
Phys. Rev. B {\bf 53}, 5670 (1996).
\bibitem{Liu} M. Liu, Phys. Rev. B {\bf 18}, 1165 (1978).
\bibitem{Martin} P.C. Martin, O. Parodi and P.S. Pershan, Phys. Rev. A
{\bf 6}, 2401 (1972).
\bibitem{Chen} M.T. Chen , J. Roesler and J.M. Mochel, J. Low Temp. Phys
{\bf 89}, 125 (1992); J.M. Mochel and M.T. Chen, Phys. B {\bf 197}, 278 (1994).
\bibitem{Kleinert3} H. Kleinert, {\it Gauge Fields in Condensed Matter
Physics}, (World Scientific, Singapore, 1989).
\bibitem{Kleinert1} H. Kleinert, J. Phys. A {\bf 19}, 1855
(1986).
\bibitem{Zippelius} A. Zippelius, B.I. Halperin and D.R. Nelson, Phys.
Rev. B {\bf 22}, 2514 (1980).
\bibitem{Hirth} J.P.Hirth and J.Lothe, {\it Theory of Dislocations}
(McGraw-Hill, New York, 1968).
\bibitem{Landau} L.D. Landau and E.M. Lifschitz, {\it Theory of Elasticity}
(Pergamon Press, Oxford, 1970).
\bibitem{Negele} J.W. Negele and Orland, {\it Quantum Many-Particle
Systems}, (Addison-Wesley, New York, 1988).
\bibitem{Hamermesh} M.Hamermesh, {\it Group Theory} (Addison-Wesley,
London, 1962) .
\bibitem{Forster} D.Forster, {\it Hydrodynamic Fluctuations, Broken Symmetry,
and Correlation Functions} (W.A. Benjamin, Reading, Massachusetts, 1975).
\bibitem{Popov} V.N. Popov, {\it Functional Integrals in Quantum Field Theory
and Statistical Physics} (Reidel, Dordrecht, 1983), Chapter 6.
\bibitem{Putterman} S.J. Putterman, {\it Superfluid Hydrodynamics},
(North-Holland, Amsterdam, 1974).
\bibitem{Jahnig} F. J\"ahnig and H. Schmidt, Ann. of Phys. {\bf 71}, 129
(1972).
\bibitem{Kirkpatrick} T.R. Kirkpatrick and J.R. Dorfman, J. Low. Temp. Phys.
{\bf 58}, 301 (1985); {\it ibid.}, {\bf 58}, 399 (1985).
\end{references}
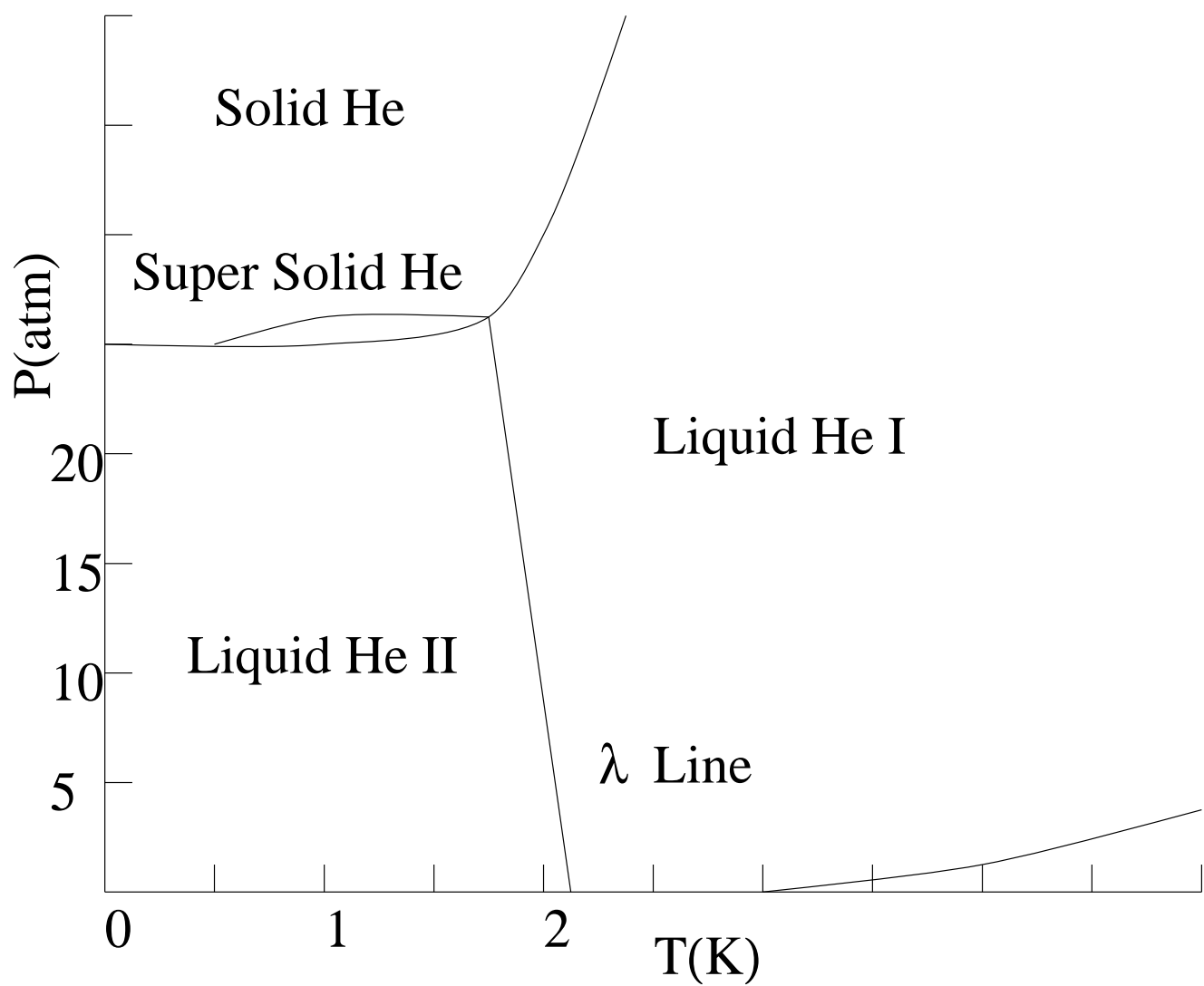
\end{document}